\documentclass{article}
\usepackage{amsmath}
\usepackage{graphicx} 
\usepackage{float}
\usepackage{authblk}
\usepackage{biblatex}         
\usepackage{booktabs}
\usepackage[utf8]{inputenc}
\usepackage{textgreek}
\title{\vspace{-4cm}Neurosymbolic Learning for Predicting Cell Fate Decisions from Longitudinal Single-Cell Transcriptomics in Paediatric Acute Myeloid Leukemia}

\author[1,2,3]{Abicumaran Uthamacumaran}
\author[1,3,4,5]{Hector Zenil\thanks{Corresponding author: hector.zenil@algocyte.ai, hector.zenil@kcl.ac.uk}}

\affil[1]{\normalsize\text{ }Oxford Immune Algorithmics, Oxford University Innovation and London Institute of Healthcare Engineering, Oxford and London, U.K.}
\affil[2]{\normalsize\text{ }McGill University, Departments of Surgical and Interventional Sciences, Neurosurgical Simulation and Artificial Intelligence Learning Centre, Montreal, Canada.}
\affil[3]{\normalsize\text{ }Algorithmic Dynamics Lab, Research Departments of Biomedical Computing and Digital Twins, School of Biomedical Engineering and Imaging Sciences, King's College London, U.K.}
\affil[4]{\normalsize\text{ }King's Institute for Artificial Intelligence, King's College London, U.K.}
\affil[5]{\normalsize\text{ }Cancer Research Interest Group, The Francis Crick Institute, London, U.K.\vspace{-1cm}}

\date{}

\begin{document}

\date{}

\maketitle

\begin{abstract}
Paediatric Acute Myeloid Leukemia (AML) is a complex adaptive ecosystem with high morbidity. Current trajectory inference algorithms struggle to predict causal dynamics in AML progression, including relapse and recurrence risk. We propose a symbolic AI and deep learning framework grounded in complexity science, integrating Recurrent Neural Networks (RNNs), Transformers, and Algorithmic Information Dynamics (AID) to model longitudinal single-cell transcriptomics and infer complex state-transitions in paediatric AML. We identify key plasticity markers as predictive signatures regulating developmental trajectories. These were derived by integrating deep learning with complex systems–based network perturbation analysis and dynamical systems theory to infer high-dimensional state-space attractors steering AML evolution.

Findings reveal dysregulated epigenetic and developmental patterning, with AML cells in maladaptive, reprogrammable plastic states, i.e., developmental arrest blocking terminal differentiation (stable equilibria). Predictions forecast neurodevelopmental and morphogenetic signatures guiding AML cell fate bifurcations, suggesting ectoderm–mesoderm cross-talk during disrupted differentiation. Neuroplasticity and neurotransmission-related transcriptional signals implicate a brain–immune–hematopoietic axis in AML cell fate cybernetics. This is the first study combining RNNs and AID to predict and decipher longitudinal patterns of cell fate transition trajectories in AML. Our complex systems approach enables causal discovery of predictive biomarkers and therapeutic targets for ecosystem engineering, cancer reversion, precision gene editing, and differentiation therapy, with strong translational potential for precision oncology and patient-centered care.\\


\end{abstract}

\section{Introduction}

Paediatric Acute Myeloid Leukemia (AML) is a complex, adaptive hematological malignancy, characterized by high morbidity, inter-patient heterogeneity, and personalized disease trajectories and therapeutic response patterns~\cite{Bolouri2018}. While the gene regulatory and molecular landscapes of pediatric AML are increasingly decoded, the dynamic network signatures that steer disease progression remain poorly understood, necessitating integrative multi-omics and advanced algorithmic approaches to resolve disease heterogeneity and predict treatment response~\cite{Edwards2025, Caplan2022, Yang2023}.  Consequently, paediatric AML recurrence remains an unmet clinical challenge, driven by emergent, maladaptive behaviors and malignant traits within AML ecologies, including therapy-induced resistance, relapse-associated clonal evolution, and tumor–microenvironmental remodeling, further compounded by the absence of predictive biomarkers. In high-risk subgroups, recurrence is fueled by multiscale reprogramming processes spanning epigenetic, transcriptomic, and metabolic interactions, that generate subtype-specific molecular heterogeneity, driving subclonal diversification and adaptive niche reshaping. These emergent dynamics remain a major barrier to effective patient care~\cite{boileau2019h3mutant, mumme2023scAML,Hanahan2022,lambo2023}.

Phenotypic plasticity—acting as the evolvability engine and cognitive scaffold of leukemic ecosystems, enables adaptive creativity, allowing cells to withstand environmental perturbations by navigating a fluid spectrum of transcriptionally, metabolically, and epigenetically reversible progenitor- or stem-like states\cite{mumme2025leukemia, dorantes2012lineage}. Thus, fate plasticity, or developmental (lineage) plasticity, enables disease progression via emergent malignant traits, such as adaptive resistance, lineage (phenotypic) switching, invasion, and stem-like behaviors without terminal commitment to a single fate or lineage \cite{Hanahan2022}. These state-transitions, or cell fate (linage) bifurcations, as a result of stalled or disrupted developmental processes, are driven by gene regulatory network dynamics~\cite{Groves2023}. To our therapeutic advantage, the resilience of these cancers—rooted in their plasticity—also marks a point of vulnerability, where targeting bifurcation signatures (plasticity markers) embedded within these regulatory networks, can enable the reprogramming of malignant cell fates. Therefore, decoding these plasticity signatures, or transition genes, regulating cell fate trajectories is critical for identifying therapeutic targets, discovering biomarkers, and improving risk stratification in precision oncology. 

While earlier models posited that leukemogenesis and progression were sustained by a discrete population of leukemic stem cells (LSCs) with self-renewal capacity, current systems-oncology views—advanced by single-cell multi-omics lineage tracing—frame AML as a complex dynamical system. In this perspective, stem-like or progenitor-like states navigate a fluid, adaptive spectrum of heterogeneous phenotypes with partial lineage commitment, trapped in stalled differentiation trajectories~\cite{Yamazaki2013Epigenome,Ling2024Aberrant}. Hematopoietic differentiation is disrupted, activating aberrant stem-like or embryonic (fetal) developmental programs that steer pathological attractor dynamics, especially in pediatric leukemias~\cite{Ling2024Aberrant}.

Here, attractors represent long-term behavioral patterns toward which transcriptional processes driving cell fate dynamics converge ~\cite{uthamacumaranzenil2022}. Reversible chromatin states and transcriptional plasticity can anchor cells at intermediate or high-potential attractor states—unstable, non-fixed positions in a plastic differentiation hierarchy. Developmental processes such as lineage bifurcations give rise to these attractor patterns; when dysregulated, they may generate chaotic or strange attractors in morphospace~\cite{uthamacumaranzenil2022,Uthamacumaran2025}. Plasticity markers in this framework correspond to bifurcation signatures—transition genes that predict state shifts across AML progression.

Conventional cell fate trajectory inference methods—such as pseudotime ordering, manifold learning, and clustering—reduce these multiscale, nonlinear dynamics to stable basins in a static multi-omic space, thereby overlooking the fluid and transient cell fate transitions that arise from the complex dynamical behaviors of aggressive subtypes~\cite{Huang2009,Li2016,uthamacumaranzenil2022}. Longitudinal single-cell studies show that relapse is driven not only by residual clones but by a hijacked microenvironment promoting immune evasion, metabolic rewiring, and stem-like plastic states~\cite{mumme2023scAML,lambo2023}. These findings demand predictive, causal inference frameworks akin to weather forecasting—able to model nonlinear interactions, bifurcations, and ecological feedbacks in tumor ecosystems.

Key molecular drivers (FLT3-ITD, NPM1, WT1, KMT2A rearrangements, fusions, dysregulated morphogenetic programs) emerge within these unstable landscapes~\cite{mumme2023scAML,lambo2023,Edwards2025,PDQ2025}. Pediatric AML exhibits age-specific drivers—\textit{KMT2A} fusions in infants, \textit{RUNX1–RUNX1T1} in older children—often lacking adult AML mutations (\textit{DNMT3A}, \textit{TP53}, \textit{IDH1/2})~\cite{Chen2024DevOrigins}. These arise in developmentally plastic hematopoietic precursors, disrupting lineage commitment and enabling hybrid cellular identities.

However, the complex network dynamics orchestrating plasticity remain poorly resolved. Identifying these regulatory programs requires dynamical systems theory to reconstruct attractor geometries and infer bifurcation signatures from longitudinal data. Plasticity signatures, i.e., transition genes or gene programs regulating reprogrammable fate decisions, demand predictive modelling and causal inference algorithms that can decode cell-fate cybernetics, forecast dysregulated trajectories, and map the network markers controlling state-transitions.

Complex systems theory, or complexity science, captures these fate decisions as emergent behaviors of complex dynamical systems (processes), shaped by bifurcations, self-organization, and causal geometries in a Waddington-like landscape~\cite{uthamacumaranzenil2022}. Deep learning models such as recurrent neural networks (RNNs) and transformers can identify latent plasticity biomarkers and temporal dependencies underlying AML recurrence~\cite{Whata2022,Xu2023,Chen2025}, but are underused due to explainability challenges. To address this, we integrated algorithmic information dynamics (AID) with network perturbation analysis via the Block Decomposition Method (BDM)~\cite{zenil2019algorithmic,Zenil2019,Zenil2018,uthamacumaranzenil2022,Uthamacumaran2022,Uthamacumaran2025} and explainable architectures (LSTMs, attention-equipped transformers) with feature-importance detection.

AID quantifies causal structure in gene regulatory networks via graph-based algorithmic complexity (Kolmogorov complexity) beyond statistical correlations~\cite{Zenil2019,zenil2019algorithmic}. Coupled with RNNs/transformers, it enables explainable prediction of cell-fate decisions from longitudinal single-cell data, reconstructing the AML transcriptional state-space, revealing attractor states, and identifying bifurcation signatures that govern disease progression. This is critical, as therapeutic resistance often arises from plasticity-driven nonlinear phase transitions near tipping points~\cite{Deb2022,Mojtahedi2016,Lee2024,Tsuchiya2016,Sarkar2019}.

Recent free-energy–based models of leukemic progression~\cite{uechi2023free,uechi2024transcriptome,frankhouser2024state} capture some transition points but may miss hidden nonlinear attractors and oscillatory, unstable states between critical points~\cite{uthamacumaranzenil2022}. By combining AID complexity measures with deep time-series models, we move beyond static Waddington-like reconstructions and statistical model-driven assumptions, to infer hidden attractor dynamics, decode bifurcation signatures, and identify the transition genes and regulatory signals steering plasticity-driven AML cell-fate transitions.

\section{Methods}

\section*{Datasets and Preprocessing}

To characterize acute AML progression, developmental plasticity, and state-transition dynamics, we integrated our complex systems approaches with bulk transcriptomic data from healthy (controls) and paediatric leukemia patients, and single-cell RNA sequencing (scRNA-seq) from patient-matched paediatric AML samples across diagnosis, remission, and relapse stages. All datasets were publicly available, previously published, and not generated by this study, as our work focuses solely on the algorithmic discovery. 

\subsection*{Healthy Control Whole Blood Transcriptome (Bulk)}
For healthy baseline comparisons, we used microarray gene expression data from whole blood RNA of five healthy individuals (GSE6351) profiled using the Affymetrix Human Genome U133 Plus 2.0 platform (GPL570). The dataset was originally published to explore disease susceptibility by blood-derived expression profiles and provides high-quality reference transcriptomes from peripheral blood mononuclear cells (PBMCs) of unaffected individuals~\cite{vahteristo2007blood}.

\subsection*{Bulk Pediatric Leukemia Transcriptome Data}
To identify differentially expressed genes (DEGs) between AML and healthy states, we used microarray gene expression data from paediatric acute leukemias (GSE7757), which included AML and other subtypes such as acute lymphoblastic leukemia (ALL). A total of 27 patient samples were profiled in this dataset~\cite{campo2007mage}. Although not limited to AML, these bulk data offer valuable insight into global transcriptional deregulation, and attractor dynamics in leukemic subtypes. We included all paediatric leukemia samples, irrespective of subtype, to capture the broader leukemogenesis landscape to compare against the healthy PBMCs, providing a bulk‐averaged view of leukemogenic signatures and collective (emergent) behaviors. This served as a complementary layer to our primary mechanistic focus of single‐cell trajectory inference in pediatric AML from diagnosis to relapse, allowing comparison of bulk leukemogenesis patterns with AML subtype‐specific dynamics in single‐cell fate decisions.

\subsection*{Longitudinal Single-Cell Transcriptomic Profiling of paediatric AML}
Our primary dataset was derived from the study by Lambo et al.~\cite{lambo2023}, which presents a longitudinal single-cell atlas of paediatric AML response to treatment. The scRNA-seq data are accessible under GEO accession \textbf{GSE235063}. Samples were collected from bone marrow or peripheral blood of 28 patients at diagnosis (Dx), remission (Rem), and relapse (Rel) timepoints. Cells were enriched via Ficoll separation, and scRNA-seq libraries were generated using the 10X Genomics platform (Chromium Single Cell 3' v3 chemistry), yielding expression profiles for 684,031 single cells. 

The authors also performed single-cell ATAC-seq on the same matched samples, allowing for multi-omics analysis of epigenetic accessibility alongside transcriptomic changes. These data reveal significant reprogramming of leukemic hierarchies toward primitive progenitor or stem-like states upon relapse, irrespective of AML subtype \cite{lambo2023}. We limited our analyses to transcriptional data to test whether our combined neurosymbolic algorithms can infer such cell fate state-transitions to primitive, or stalled developmental programs without the additional multiomics input. Due to computational constraints, we randomly selected (half) 14 of the 28 patient trajectories spanning all three clinical timepoints (Dx, Rel, Rem) for predictive modelling and causal inference analysis. These samples represent heterogeneous AML subtypes and were chosen to preserve both subtype diversity and trajectory completeness. 

\subsection*{Rationale for Dataset Integration}
This integrated design enabled multi-resolution modelling of AML dynamics. Bulk transcriptomes provided group-level trends and robust DEG discovery for disease vs. control comparisons, while scRNA-seq enabled high-fidelity resolution of intratumoral heterogeneity and temporal shifts in cellular hierarchies. The combination of healthy blood, leukemic bulk RNA, and longitudinal single-cell AML profiles allowed us to extract candidate plasticity markers and validate them across resolution scales.

\subsection*{DEG Extraction and Statistical Analysis}
Differential expression analysis (DEG) was performed on the bulk RNA and scRNA-seq data. For the bulk RNA microarray datasets of acute paediatric leukemias and healthy controls (raw \texttt{.CEL} files) were normalized using Robust Multi-array Average (RMA) via the \texttt{affy} R-package. Samples were grouped into treatment (T) and normal (N) cohorts based on GEO accession metadata. Probe identifiers were mapped to gene symbols and Entrez IDs using Bioconductor, retaining only unique mappings, and the matrix was subjected to downstream network analyses. A design matrix was constructed, and linear modelling was applied using the \texttt{limma} package. Empirical Bayes moderation was used to shrink standard errors and improve statistical power. Differentially expressed genes were identified using the \texttt{topTable} function with Benjamini–Hochberg adjustment for multiple testing.   
For the single-cell data, DEG was performed across three temporal states of gene expression: DX, REL, and REM. For each gene, pairwise $log_2$ fold changes (\textit{LogFC}) were calculated between the three states. Two-sample \textit{t}-tests were conducted for each comparison, with p-values adjusted using the Benjamini-Hochberg method (False Discovery Rate, FDR). Genes with adjusted p-values ($p_\text{adj} < 0.05$) and $\lvert\log_2\text{FoldChange}\rvert \geq 1$ are considered significant. From these, the top 100 DEGs, ranked by absolute \textit{LogFC}, were selected for downstream analysis to capture longitudinal state-transition dynamics, and predict plasticity regulators steering the cell fate trajectories. Normalized gene expression datasets were used for all analyses, with a focus on protein-coding gene transcripts across all analyses. g:Profiler, a web-based gene set enrichment tool, was used to functionally annotate key differentially expressed gene signatures. For biological relevance and interpretation, gene functions and disease associations were cross-referenced using the online GeneCards database, with key findings discussed in the results.

\subsection*{BDM Analysis}

BDM analysis was applied to quantify the algorithmic complexity of gene regulatory networks inferred from expression data. Using the \texttt{pyBDM} Python package, adjacency matrices were first constructed from the Spearman correlation coefficients between DEGs. These correlation matrices were binarized using a threshold of 0.5 to yield undirected, unweighted graphs. The \texttt{pyBDM} package approximates algorithmic complexity by decomposing each adjacency matrix into smaller overlapping square blocks, estimating the algorithmic probability of each block based on a precomputed lookup table of Turing machine outputs, and summing their information content with logarithmic penalization for repeated patterns.

Two complementary perturbation analyses were conducted: (1) \textbf{Node-based perturbation}, where each node (gene) was systematically zeroed out (along with its edges), and the resulting change in BDM complexity was recorded, and (2) \textbf{Link-based perturbation}, where individual edges between nodes were removed to assess the contribution of specific gene-gene interactions to the overall network structure. Perturbation scores reflect the drop (or rise) in algorithmic complexity, capturing both structural and causal relationships of nodes and links in the network, beyond what entropy or centrality measures-based network metrics can reveal.

\subsection*{NNMF Analysis}
Non-negative Matrix Factorization (NNMF) is applied as an unsupervised hierarchical clustering algorithm (Modularity detection) to decompose the gene expression matrices into transcriptional co-expression modules. The number of components was set to 5, and for each module, the top 50 genes were identified based on their module scores. These genes were used to construct Spearman adjacency matrices, followed by binarization at a threshold of 0.5. Both node-based and link-based BDM perturbation analyses were performed on these networks. Ribosomal genes and non-coding signatures were filtered out for functional marker enrichment analyses using g:Profiler.

\subsection*{LSTM Analysis}
A Long Short-Term Memory (LSTM) network was utilized to analyze temporal dependencies in longitudinal gene expression patterns across DX, REL, and REM states. The LSTM and BiLSTM models were implemented using the \texttt{torch.nn.LSTM} module from the PyTorch framework, an open-source deep learning library for Python. Gene expression data was normalized using the MinMaxScaler, reshaped to $(samples, timesteps,\\ features)$, and split into training (70\%), validation (15\%), and testing (15\%) datasets. The LSTM architecture consisted of a single LSTM layer with 64 units and a Dense output layer with a softmax activation for multi-class classification. The model was compiled using the Adam optimizer and sparse categorical cross-entropy loss and trained for 50 epochs with a batch size of 16. Feature importance was computed from the LSTM input weights matrix by taking the mean absolute value across gates/units. The top 100 genes ranked by feature importance were extracted and subjected to BDM perturbation analysis.

\subsubsection*{Bidirectional LSTM} The Bidirectional LSTM (BiLSTM) model comprises two bidirectional LSTM layers: the first layer with 64 units outputs sequences for the next layer, and the second layer with 32 units outputs a single sequence. Dropout layers are applied after each LSTM layer to prevent overfitting, with dropout rates set at 30\% and 20\%. Dense layers include a fully connected layer with 32 units using ReLU activation, followed by a final softmax layer for three-class classification. The model is trained using the Adam optimizer, sparse categorical cross-entropy loss, 50 epochs, and a batch size of 16. This architecture enables the extraction of sequence-based features and robust classification for patient-specific data. Gene (feature)importance was computed directly from the learned weights. 

\subsection*{Transformer Analysis}
A Transformer model was utilized to analyze temporal gene expression patterns across the three time-points: DX, REL, and REM states. The transformer was implemented using the \texttt{torch.nn.Transformer} module from PyTorch. The transformer architecture included a multi-head self-attention mechanism with 4 attention heads, a hidden dimension matching the input features, and a feedforward neural network comprising two Dense layers with 64 and input-matching dimensions, respectively. Layer normalization, dropout (rate = 0.1), and residual connections were applied for stability and regularization. Gene expression data was normalized using MinMaxScaler, reshaped to $(samples, timesteps, features)$, and subjected to a 70:15:15 split for training:testing:validation. The model was trained using the Adam optimizer, sparse categorical cross-entropy loss, and a batch size of 16 for 50 epochs. Feature importance was derived from the MultiHeadAttention outputs layer by extracting attention weights and averaging their absolute values across heads and timesteps. The top 100 genes ranked by importance scores were obtained. 

\begin{figure}[H]
    \centering
    \includegraphics[width=0.9\textwidth]{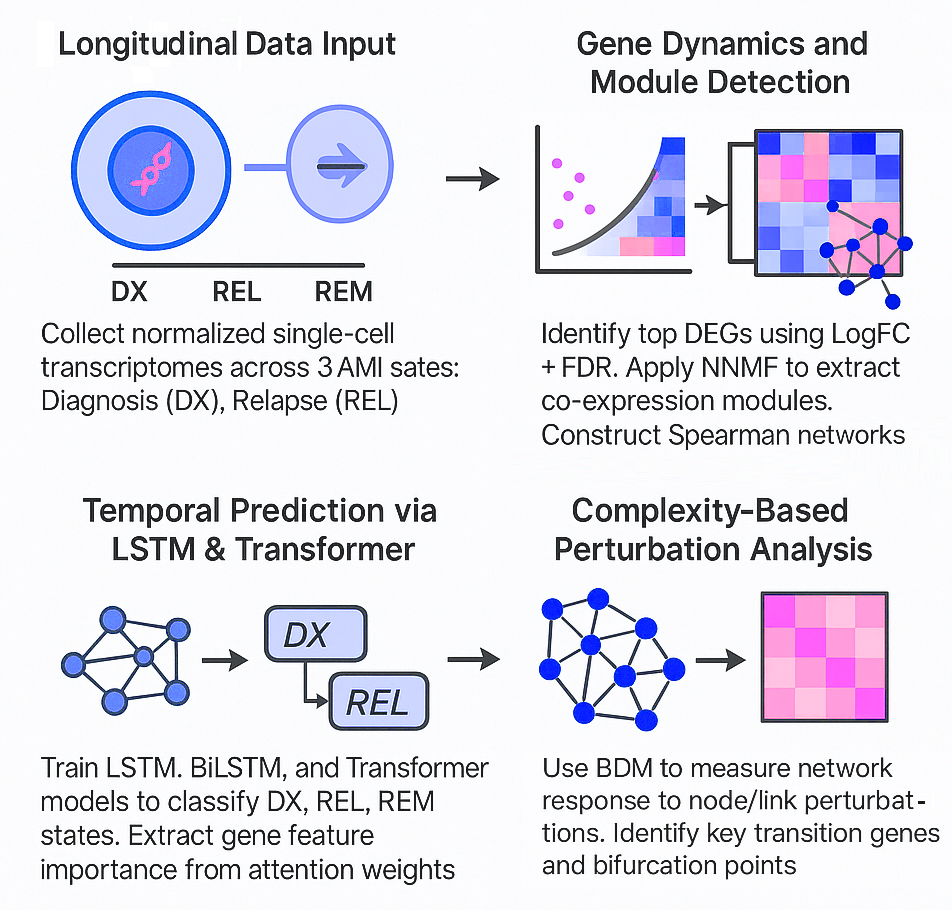}
    \caption{\textbf{Symbolic deep learning algorithms to predict AML state-transitions and deconstruct longitudinal single-cell transcriptional landscapes.}  
    The schematic outlines a four-step workflow starting from single-cell gene expression matrices across diagnosis (DX), relapse (REL), and remission (REM) AML states. It integrates differential gene and module detection, followed by deep learning-based temporal classification, and BDM-based perturbation analysis to identify critical transition genes steering cell fate bifurcations in paediatric AML trajectories.}
    \label{fig:aml_symbolic_workflow}
\end{figure}

\subsection*{Bayesian Ridge Regression}

Bayesian Ridge Regression (BRR) is a form of regression algorithm using a probabilistic Bayesian framework that estimates posterior distributions of model parameters, making it robust to small sample sizes and noise in longitudinal gene expression. BRR was used as a comparison of causal inference tool with AID approaches. This allows causal inference by incorporating uncertainty in coefficient estimates, for predictive modelling of longitudinal transcriptional dynamics and causal inference throughout disease progression. BRR was implemented using \texttt{sklearn.linear\_model.BayesianRidge} to model longitudinal trajectories. The input consisted of gene expression matrices. For each patient sample, expression data was preprocessed, normalized, and analyzed by the BRR model. The hyperparameters used included $\alpha_1=1e^{-6}$, $\alpha_2=1e^{-6}$ for weight precision priors, $\lambda_1=1e^{-6}$, $\lambda_2=1e^{-6}$ for noise precision priors, a tolerance of $10^{-4}$, and 300 iterations.

\section{Results}

\subsection*{Conserved but Reprogrammed Epigenetic and Chromatin-Linked Co-expression Modules in Paediatric Acute Leukemias}

\begin{figure}[H]
    \centering
    \hspace*{-1.5cm} 
    \includegraphics[width=1.1\textwidth]{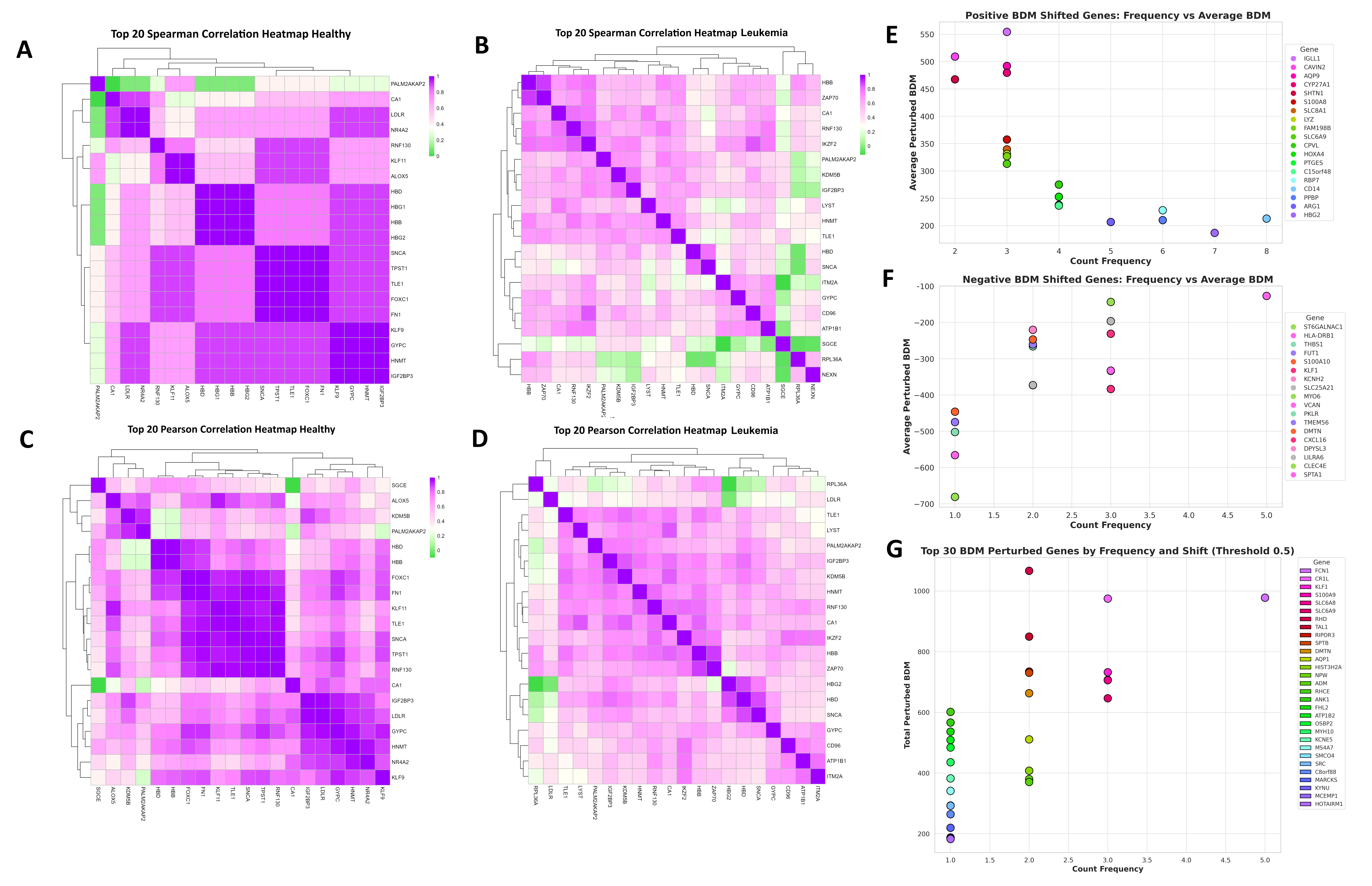}
    \caption{Top 20 Spearman and Pearson Correlation Heatmaps for Healthy and paediatric Acute Leukemia Conditions. 
    \textbf{(A–D) Bulk Leukemia vs Healthy PBMC Bulk RNA:} (A) Spearman correlation in healthy samples, (B) Spearman correlation in leukemia samples, 
    (C) Pearson correlation in healthy samples, (D) Pearson correlation in leukemia samples. 
    \textbf{(E–G) Single‐cell AML Subtype Only:} (E) Positive and (F) Negative BDM‐shifted genes: count frequency vs average BDM perturbation. Circle size and color indicate unique gene identities, with larger circle edge widths for visibility. (G) Top BDM‐perturbed genes ranked by count frequency and total BDM shift using a binary threshold of 0.5. \textit{Note: Panels E–G refer specifically to AML subtypes only, in contrast to panels A–D, which compare all paediatric acute leukemias with healthy PBMCs as a reference for attractor dynamics and developmental plasticity.}}
    \label{fig:bulk_aml_heatmap}
\end{figure}

To characterize the broader transcriptional landscape of paediatric acute leukemias, we first performed correlation network analysis on the top DEGs (by highest log-fold change, and statistical significance) from bulk RNA data. This approach highlights global co-expression patterns across healthy and leukemic samples, providing a macroscopic baseline of regulatory architecture. These correlation structures offer a useful contrast to the fine-grained, temporally resolved dynamics inferred from single-cell data, revealing how cell fate bifurcations in acute leukemias may diverge from the broader paediatric leukemia developmental landscape. Figure~\ref{fig:bulk_aml_heatmap} presents Spearman and Pearson correlation heatmaps of the top 20 DEGs across healthy blood and various paediatric acute leukemia subtypes. This global bulk RNA landscape provides a macroscopic view of co-expression patterns, serving as a reference against which the dynamic, and potentially divergent cell fate trajectories observed in longitudinal AML single-cell data can be compared to identify bifurcations and lineage-specific transitions.

Strong correlations (above $|0.6|$) are observed among genes such as \textit{HBB}, \textit{ZNF130}, \textit{FOXC1}, and \textit{PALM2AKAP2}, indicating their coordinated expression patterns in both healthy and AML states. Notably, these correlations shift between conditions, suggesting reprogramming of regulatory networks in AML.

Several genes in these correlations have key epigenetic roles, as revealed by GeneCards Dabatase. For instance, FOXC1, a forkhead transcription factor, is crucial in hematopoiesis and mesenchymal differentiation (or EMT like plastic transitions). TLE1, a Groucho family member, participates in developmental pathways via histone deacetylase interactions. CD96 (TACTILE), an immunoglobulin-like receptor, is consistently correlated in both Pearson and Spearman analyses, highlighting its role in immune evasion and AML progression. Similarly, KDM5B (Lysine Demethylase 5B), a histone demethylase regulating H3K4 methylation, was observed as a central driver of cell fate differentiation in our previous study on single-cell pHGG (paediatric high-grade glioma) dynamics~\cite{Uthamacumaran2022,Uthamacumaran2025}. Its presence here reinforces its role in stemness, plasticity, and differentiation control in aggressive paediatric cancers. Moreover, HNMT (Histamine N-methyltransferase), a key enzyme in histamine metabolism, is another gene commonly correlated in both Spearman and Pearson heatmaps. 

\begin{table}[H]
    \centering
    \caption{Genes with Epigenetic and Developmental Roles in Leukemia and Healthy Samples}
    \label{tab:epigenetic_genes}
    \begin{tabular}{ll}
        \toprule
        \textbf{Gene} & \textbf{Function} \\
        \midrule
        CD96 & Immune checkpoint regulation, leukemia progression \\
        KDM5B & H3K4 demethylation, cell fate differentiation \\
        HNMT & Histamine metabolism, immune modulation \\
        FOXC1 & Hematopoiesis, differentiation \\
        TLE1 & Chromatin remodelling, transcriptional repression \\
        \bottomrule
    \end{tabular}
\end{table}

These major findings suggest a chromatin remodelling underlying the state-transitions from healthy to malignant (leukemic) transcriptional regulation of hematopoietic systems,  driven by complex microenvironmental-immune interactions. The presence of CD96, FOXC1, KDM5B, and HNMT in both correlation analyses further suggests their epigenetic role in leukemogenesis, and leukemic dynamics and aligns with our prior observations in paediatric glioma, suggesting common mechanisms of differentiation control and tumour evolution of common paediatric cancers. Several of the identified genes, such as KDM5B and TLE1, are strongly implicated in developmental patterning processes regulated by WNT signaling and histone methylation dynamics. Specifically, KDM5B modulates H3K4me3 levels to repress WNT target genes, while TLE1 acts as a canonical WNT repressor, both contributing to chromatin remodeling during neurodevelopment and lineage specification (i.e., a shared neurodevelopmental axis). Hence, we propose that modulating H3K4me3 levels and the WNT signaling pathway could serve as a `differentiation therapy' or lineage (fate) 'transition therapy'~\cite{aguade2022transition}, to control leukemia aggressiveness (progression) or promote cancer reversion to healthy-like behavioural states by restoring developmental patterning and differentiation potential. We suggest that targeted differentiation therapies via our identified plasticity signatures could redirect malignant cell states, and their maladaptive behaviors toward more stable, non-invasive fates.
 
\subsection*{Single-Cell BDM Perturbation Signatures Reveal Plasticity Regulators and Lineage Cross-talk Driving AML State-Transitions}

To move beyond the averaged signals of bulk leukemic transcriptomes and resolve the transcriptional heterogeneity underlying cell fate decisions in the AML subtype, we dissected longitudinal single-cell gene expression patterns. For the single-cell RNA-seq data, we employed a pseudobulk differential expression approach across the DX, REL, and REM states. This method captures temporal changes and collective behaviors in single-cell gene expression patterns while preserving cell-to-cell variability, offering finer resolution of state-specific DEGs than traditional bulk RNA-seq. We used this simpler approach as a proxy baseline to compare against deep learning–selected feature importance genes, aiming to assess overlap in predictive biomarkers and enhance interpretability by integrating BDM’s causal inference for robust predictive signature discovery.

Figure~\ref{fig:bulk_aml_heatmap}, panel E highlights the top DEGs with the highest positive perturbed BDM shifts across the AML scRNA-Seq data with a 0.1 binarization threshold, where IGLL1 (554.3) and AQP9 (492.0) exhibit the most significant shifts. According to GeneCard Database, these genes play crucial roles in immune system regulation and cellular transport. 

The presence of S100A8 (357.5) and ARG1 (206.7) further indicates a shift in immune-inflammatory and metabolic pathways. g:Profiler gene set enrichment identified secretory vesicles and granules as central functions of these genes, as well as LYZ and CD14 as macrophage or microglial markers. Further, the transcription factors TEF and NFIL3 were found through enrichment analysis, which are involved in embryonic development, neuro-endocrine signaling, and immune regulation. Among the top genes exhibiting the highest positive perturbed BDM shifts, several key immune regulatory markers emerged with potential therapeutic relevance. RBP7 encodes a retinol-binding protein involved in vitamin A transport and signaling, linking BDM perturbation to metabolic and differentiation pathways in glioma.  RBP7 may serve as a promising target for metabolic reprogramming and differentiation therapy. For instance, retinoids—particularly retinoic acid, a metabolite of Vitamin A—have been shown to induce neuroblastoma cell differentiation into neuron-like states, reducing malignancy~\cite{bayeva2021retinoic,illendula2020retinoic}.

CD14, the most frequently perturbed gene across the 14 AML patient samples, encodes a pattern recognition receptor expressed on innate immune system cells such as monocytes and macrophages. Thus, targeting CD14-mediated feedback circuits may enable precise control of immune activation in AML and serve as an immunotherapeutic target against AML's niche construction. 

Additional highly perturbed genes shown here include HOXA4, a homeobox transcription factor involved in hematopoietic lineage commitment and immune cell differentiation, whose dysregulation is associated with leukemogenesis and stem cell remodeling~\cite{alharbi2013role,elias2018hoxa}. We propose these BDM signatures may serve as targeted therapies for AML state-transition control to modulate tumour microenvironments and guide cell fate-immune reprogramming.

Figure~\ref{fig:bulk_aml_heatmap} panel F presents the top DEGs with the most negative perturbed BDM shifts with a 0.1 binary threshold, with ST6GALNAC1 (-681.1) and HLA-DRB1 (-565.9) showing the largest shifts, implicating immune modulation and glycosylation disruptions. KLF1 (-383.6), a key regulator of erythroid differentiation, suggests hematopoietic perturbations, while S100A10 (-446.1) and THBS1 (-502.2) reflect changes in tumour microenvironment remodeling and angiogenesis.

KLF1, VCAN, and MYO6 also exhibited strong negative shifts and are known to regulate erythroid differentiation, extracellular matrix (ECM) remodelling interactions, and cytoskeletal transport—processes essential for maintaining stromal integrity and hematopoietic signaling. Other ECM and adhesion remodeling-associated signatures include VCAN, and THBS1. Negative BDM perturbations of these genes  may reflect a collapse of bioelectrical and mechanical transduction circuits that normally constrain leukemic cell fate transitions. In evidence, among the observed negative BDM sigantures, KCNH2 encodes a voltage-gated potassium channel (hERG1) critical for maintaining cellular membrane potential and bioelectric homeostasis. S100A10, a calcium-binding protein, modulates membrane repair and ion transport, contributing to cellular excitability and signaling plasticity. Hence, these BDM perturbation signatures likely denote plasticity regulators regulating AML state-transitions and behavioral patterns towards aggressivity. 

Figure~\ref{fig:bulk_aml_heatmap}, panel G highlights the top DEGs by BDM perturbation analysis using a 0.5 binarization threshold, a less granular approach, whereas a 0.1 threshold captures more subtle shifts, into network plasticity and resilience in cell fate transition dynamics. As seen, RHD (1066.3), CR1L (974.7), and FCN1 (977.7) show the highest BDM perturbed shifts. These genes are involved in erythropoiesis, complement system regulation, and immune response, indicating significant adaptive niche construction and tumor-immune microenvironmental reshaping. The presence of KLF1 (732.1) and S100A9 (706.9) further supports a hematopoietic and inflammatory remodelling responses, highlighting the reorganization of regulatory networks in response to selective pressures. SRC, for instance, is involved in focal adhesion dynamics of cells and their ECM remodelling during invasion. g:Profiler analysis revealed ammonium channel activity, ankyrin-1 complex activity, and CAMK2 pathways as significant enrichments. 

Several of the top perturbed BDM signatures, including SLC6A8, SLC6A9, and KCNE5, are solute carriers and ion channel (bioelectric signaling) regulators integral to maintaining cellular electrochemical gradients and membrane excitability. For instance, SLC6 transporters mediate neurotransmitter transport, linking metabolic signaling with bioelectric states. KCNE5 encodes a potassium channel subunit influencing cellular repolarization and membrane plasticity. The significant shifts in these transport and membrane potential regulating genes suggest altered bioelectric signaling and ion flow, which may dysregulate collective behaviors, i.e., cell-cell and cell-environment communication networks, and thereby, destabilize cellular identity, contributing to AML cell fate reprogramming and plasticity. This supports the view of cancers such as AML as disorders of stalled differentiation and developmental plasticity, reflecting disrupted lineage identity and aberrant cell fate trajectories.

Table~\ref{tab:nnmf_bdm_table} presents the top gene signatures exhibiting the most significant BDM  shifts identified via NNMF clustering of differentially expressed genes. These signatures reveal critical regulators of cell fate bifurcations, developmental dynamics, and tumor-immune microenvironment plasticity. Among the most negatively shifted genes are NACA (-405.7), EEF1A1 (-378.3), ACTG1 (-363.0), TPT1 (-323.3), and H3F3B (-257.5), whose downregulation suggests destabilized differentiation processes in leukemic networks. Further, S100A4 (-312.9) and VIM (-190.3) may reflect altered critical state-transitions between quiescent stem-like and invasive states, or hematopoietic-to-mesenchymal plasticity underlying leukemic dissemination and therapy resistance. HMGB1 (-200.6) and HLA-DRA (-241.1) point further to dysregulated immune-inflammatory signaling and plasticity markers. Conversely, the highest positive BDM shifts in IGKC (265.4) and IGLC2 (25.4) suggest hijacking of B-cell (lymphoid) lineage differentiation programs, and upregulation of S100A9 (74.6) implies calcium-mediated immune remodeling and inflammatory niche construction. These findings are consistent with calcium ion binding emerging as the most statistically significant gene set enrichment from Bayesian Ridge regression (Figure 3G). Collectively, these complexity-derived plasticity regulators suggest causal attractor transitions wherein AML cell fate plasticity seems to be situated at the interface of neurodevelopmental and hematopoietic differentiation programs, highlighting ectoderm–mesoderm cross-talk steering AML cell fate bifurcations.

Furthermore, the presence of lymphoid transcriptional programs in AML single-cell BDM signatures suggests a hybrid, and adaptive fluid continuum of phenotypes, and a primitive, mixed-lineage origin or developmental arrest, where AML leukemic cells retain features of early hematopoietic progenitors. Lambo et al. (2023) demonstrated through multiomics analyses that such leukemias often exhibit lymphoid-like chromatin accessibility and transcriptional signatures, despite being phenotypically myeloid~\cite{lambo2023}. This highlights a shared progenitor state and suggests lineage plasticity may underlie therapeutic resistance and relapse.

Figure~\ref{fig:bulk_aml_heatmap} Panel E shows the top genes with highest average positive BDM perturbation scores (e.g., \textit{IGLL1}, \textit{AQP9}, \textit{CAVIN2}, \textit{CYP27A1}) and their recurrence across contexts (count frequency). Panel F highlights top negative BDM-shifted genes, such as \textit{ST6GALNAC1}, \textit{HLA-DRB1}, \textit{KCNH2}, and \textit{VCAN}, implicating roles in immune modulation, ion transport, and ECM-based niche construction. These results support the use of BDM-based perturbation scoring as a sensitive network-level biomarker for identifying causal and regulatory fragility across glioma conditions. Panel G shows gene signatures such as \textit{FCN1}, \textit{CR1L}, and \textit{RHD} show the strongest perturbation impact across multiple patient samples (Count frequency). This scatter plot integrates both recurrence and perturbation strength to prioritize candidate genes.

\begin{table}[H]
\centering
\caption{BDM perturbations of NNMF-Inferred Bifurcation Signatures.}
\label{tab:nnmf_bdm_table}
\begin{tabular}{l|r}
\textbf{Gene} & \textbf{BDM Shift} \\
\hline
IGKC & 265.4 \\
SSR4 & 153.5 \\
HIST1H4C & 95.9 \\
CA2 & 88.3 \\
S100A9 & 74.6 \\
SLC25A37 & 60.6 \\
PRDX2 & 39.8 \\
GYPC & 29.0 \\
BLVRB & 26.8 \\
IGLC2 & 25.4 \\
SEC11C & 20.8 \\
SNCA & 16.0 \\
TUBA1B & 1.2 \\
HBB & 0.4 \\
GLRX5 & -145.3 \\
FTL & -153.3 \\
HBA2 & -153.6 \\
HLA-B & -158.2 \\
CA1 & -166.0 \\
TMSB4X & -171.5 \\
HBA1 & -175.5 \\
B2M & -176.7 \\
VIM & -190.3 \\
SAT1 & -197.4 \\
FAU & -197.5 \\
HMGB1 & -200.6 \\
CST3 & -211.3 \\
NEAT1 & -212.6 \\
HSP90AA1 & -234.8 \\
HLA-DRA & -241.1 \\
GYPB & -244.5 \\
UBC & -255.2 \\
H3F3B & -257.5 \\
EIF1 & -261.4 \\
LGALS1 & -268.2 \\
PTMA & -277.8 \\
SRGN & -278.2 \\
CYBA & -293.0 \\
GAPDH & -306.3 \\
S100A4 & -312.9 \\
TPT1 & -323.3 \\
ACTG1 & -363.0 \\
EEF1A1 & -378.3 \\
NACA & -405.7 \\
\end{tabular}
\end{table}

\clearpage

\subsection*{AI-Driven Longitudinal Biomarker Discovery}

\subsubsection*{Deep learning Models Reveal Temporally Predictive and Architecture-Specific Plasticity Signatures in AML Cell Fate trajectories}

\begin{figure}[H]
    \centering
    \includegraphics[width=\textwidth]{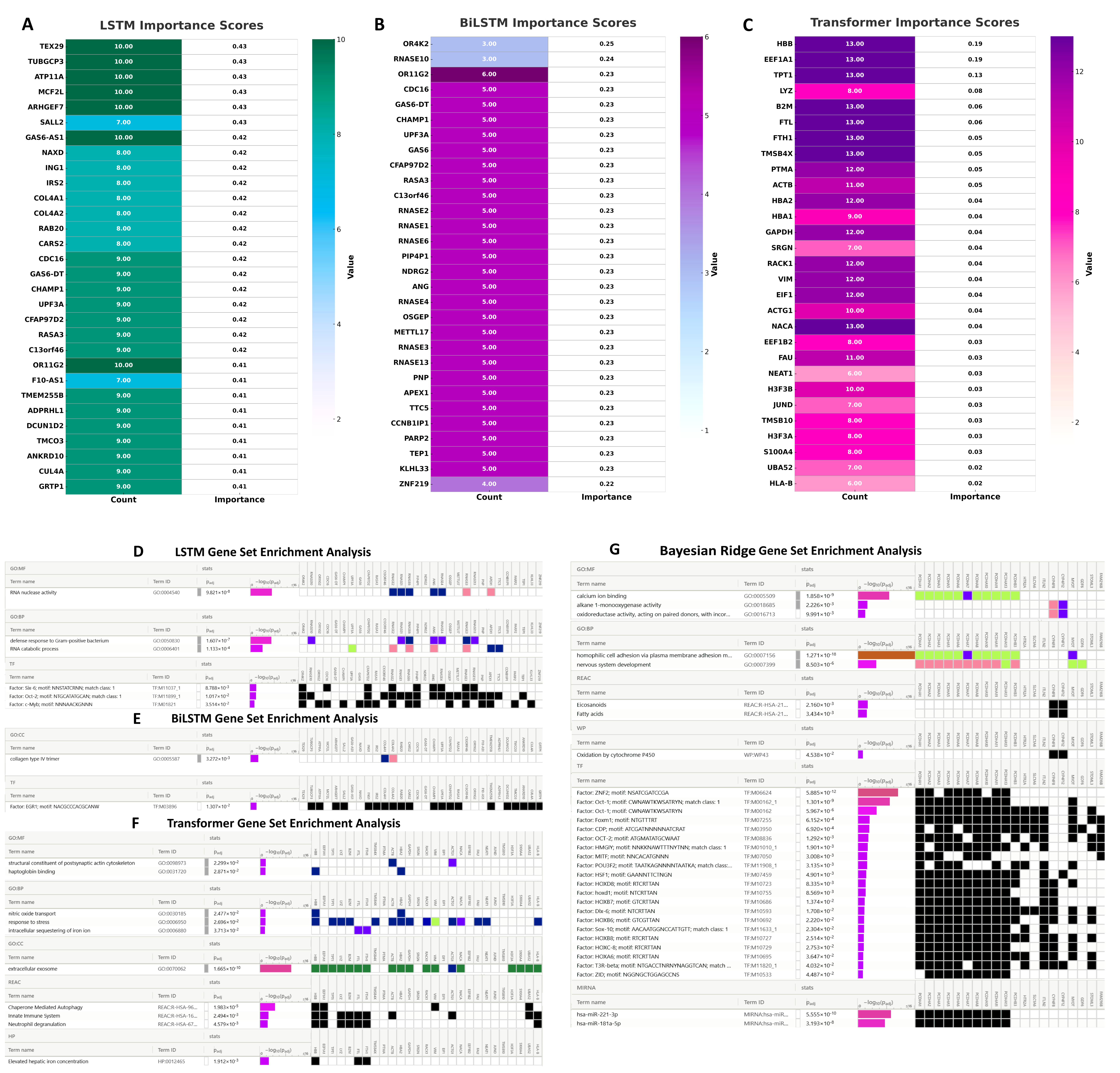}
    \caption{
    Panels A–G illustrate the top 30 genes with the highest importance scores (predictive signatures) as identified by three distinct deep learning architectures applied to longitudinal single-cell gene expression matrices of AML patient trajectories: Panel A (LSTM), Panel B (BiLSTM), and Panel C (Transformer). Gene set enrichment analysis of top-ranking genes identified by each model: LSTM (D), BiLSTM (E), Transformer (F), and Bayesian Ridge Regression (G).
    }
 \label{fig:aml_dl_importance_heatmaps}
\end{figure}

These algorithms capture nonlinear dependencies across time and feature hierarchies—LSTM (Long Short-Term Memory) and BiLSTM (Bidirectional LSTM) focus on sequence-based recurrent layers, while Transformer uses attention mechanisms across the gene expression feature space. The frequency counts indicate how many of the 14 patient samples included each gene in their top-ranked feature set. The average importance scores were derived from each model’s internal weight distribution, representing the contribution of each predictive gene to the disease trajectory. These were extracted directly from the learned patterns across architecture layers.

Panel A (LSTM): TEX29, TUBGCP3, and ATP11A emerged as highly consistent across all samples. Panel B (BiLSTM): RNASE family genes dominated with moderate importance. Panel C (Transformer): HBB, EEF1A1, and TPT1 stood out with higher significance scores and recurrence across nearly all samples. These findings highlight distinct gene sets captured by different learning paradigms, informing future multimodal feature fusion or ensemble methods for AML trajectory modeling.

\subsubsection*{Rac/Ras GTPase Signaling Genes Emerge as Drivers of Cytoskeletal-Matrix Remodeling and Leukemic Cell-State Transitions}

The top predictive genes identified by LSTM in Figure~\ref{fig:aml_dl_importance_heatmaps} panel A highlight key regulators of cytoskeletal remodeling, membrane plasticity, and intracellular signaling across the AML disease trajectory. Specifically, ARHGEF7, RAB20, and RASA3 are involved in Rho and Ras GTPase signaling pathways, which govern actin reorganization, vesicle trafficking, and cell migration—critical features of leukemic plasticity and immune evasion. The prominence of these genes suggests that Rac/Ras-driven programs underlie aberrant cell-state transitions and may serve as actionable biomarkers for differentiation and therapy resistance in paediatric AML.

In panel D of g:Profiler analysis shown in Figure~\ref{fig:aml_dl_importance_heatmaps}, LSTM-selected genes were enriched for RNA nuclease activity, defense response to Gram-positive bacterium , and RNA catabolic processes. Transcription factor (TF) motif enrichment included SIX6, a regulator of eye development; c-Myb, a proto-oncogene involved in proliferation and differentiation; and PARP, important in DNA repair. Additional enrichments included METTL17 (iron-sulfur cluster methylation), NDRG2 (stress response), and RASA3 (Ras signaling). In panel E , BiLSTM-identified markers showed significant enrichment for collagen type IV trimer  and EGR1, a TF involved in neurogenesis and differentiation. 

\clearpage

\subsubsection*{Epigenetic Reprogramming via Oncohistone Variants Dominates AML State-Transition Dynamics and Behavioral Patterns}

The Transformer model identified several epigenetically relevant genes among its top predictive features (Figure~\ref{fig:aml_dl_importance_heatmaps}, panel C). Notably, TPT1 and H3F3A/B are recognized modulators of chromatin structure and gene expression. H3F3A and H3F3B encode histone H3.3 variants known to harbor driver mutations in various cancers, including AML and paediatric high-grade gliomas~\cite{Wang2025,Jessa2019}. These oncohistones disrupt methylation patterns—particularly H3K27me3—leading to stem cell expansion, impaired differentiation, and increased leukemic aggressiveness as shown in Boileau et al. (2019)~\cite{boileau2019h3mutant}. Their presence in our top Transformer-ranked genes suggests epigenetic reprogramming is a dominant feature of longitudinal AML trajectory. Further, it demonstrates the predictive power of transformers in capturing these complex state-transition dynamics. 

As shown by g:Profiler analysis in Figure~\ref{fig:aml_dl_importance_heatmaps} panel F, Transformer-derived genes were enriched for iron-related processes including nitric oxide transport, response to stress, and intracellular iron sequestration, as well as niche construction processes and structural components like the actin cytoskeleton and haptoglobin binding. Pathways related to extracellular exosomes, chaperone-mediated autophagy, innate immune signaling, and neutrophil degranulation were also significantly enriched suggesting immune-tumor microenvironmental signaling. Additionally, elevated hepatic iron concentration  and VEGF signaling were captured among the Transformer features, consistent with NNMF pathway clustering.

\subsubsection*{Neuronal Differentiation Programs Indicate Lineage Bifurcation in AML}

Gene set enrichment analysis of top importance markers identified by Bayesian Ridge Regression are shown in Figure~\ref{fig:aml_dl_importance_heatmaps}, panel G). The analysis highlights significant enrichment in nervous system development (GO:0007399), and transcription factors enriched in neuronal differentiation processes, suggesting key roles in oncofetal development, neurogenesis and cell fate bifurcations, with possible evidence of ectoderm–mesoderm cross-talk where neuronal-like and hematopoietic identities intersect. This points to a developmental blockade or dysregulation in critical patterning processes that normally coordinate neurogenesis and hematopoiesis, whose disruption may drive AML plasticity.

\section*{Discussion}

By applying deep learning algorithms integrated with AID, we reconstructed the state‐transition dynamics of AML across disease progression and identified the causal drivers and regulators of fate plasticity. Across all deep learning models (LSTM, BiLSTM, Transformer) combined with BDM perturbation and NNMF clustering, our analyses converged on three overarching mechanisms and behavioral patterns in AML cell fate cybernetics: (i) epigenetic dysregulation (e.g., \textit{H3F3A/B}, \textit{KDM5B}, \textit{NEAT1}) driving stalled differentiation, (ii) immune reprogramming and inflammatory signaling (e.g., \textit{S100A9}, MHC class II genes) as plasticity regulators of AML ecologies, and (iii) Morphogenetic and neurodevelopmental program hijacking in AML ecologies (e.g., \textit{DLX6}, \textit{SOX10}, HOX clusters) underpinning lineage bifurcations and plasticity, with the latter suggesting that ectoderm–mesoderm crosstalk and oncofetal patterning processes may regulate these complex attractor dynamics and cell fate cybernetics.

While LSTM/BiLSTM emphasized immune-related and  bidirectional differentiation signals involved in active niche construction, the Transformer captured broader metabolic, hematopoietic, and chromatin-regulatory programs steering cell fate biases. BDM-based network perturbation analysis highlighted the causal control points (i.e., critical tipping points) steering cell fate bifurcations. Together, these complementary and hybrid approaches reconstructed the attractor geometry of AML progression, pinpointing causal transition genes as predictive biomarkers of fate plasticity and relapse risk.

\subsection*{Recurrent Neural Networks Decode Epigenetic and Developmental Plasticity Programs in AML}

This study demonstrates for the first time our understanding of AML recurrence and underlying plasticity markers using deep learning-driven, time-series forecasting methods, coupling causal inference algorithms such as BDM perturbation analysis, and RNNs. Feature importance is extracted from deep learning neural network weights, providing interpretable, explainable AI–driven biomarkers that can guide precision medicine and be tailored toward targeted therapies upon functional assay validation.

The comparative feature importance analysis of LSTM, BiLSTM, and Transformer models in longitudinal AML trajectory prediction reveals key differences in gene importance and functional regulation (Figure~\ref{fig:aml_dl_importance_heatmaps}). LSTM feature importance identifies highly ranked genes such as SALL2, TEX29, TUBGCP3, ATP11A, and MCF2L, which are involved in microtubule organization and cytoskeletal regulation, and matrix interactions crucial for developmental differentiation and fate plasticity. Genes like SALL2, ING1, and IRS2 are known regulators of leukemic cell cycle control (proliferation), apoptosis. SALL2 also encodes a transcription factor involved in embryonic developmental patterning and neuronal differentiation processes. Additionally, SALL2 and ARHGEF7 highlight transcriptional regulation and Rho-GTPase signaling, reflecting stem cell-like plasticity regulation underlying AML progression (\ref{fig:aml_dl_importance_heatmaps}). Additionally, the emergence of RAB20 and RASA3, suggest some Rac-Ras programs underlying the phenotypic plasticity, and fluid transitions. 

BiLSTM feature importance patterns (Figure\ref{fig:aml_dl_importance_heatmaps}) overlap significantly with LSTM, particularly in genes such as GAS6-DT, CHAMP1, CDC16, and RASA3. However, BiLSTM uniquely enriches RNASE family genes, suggesting a coordinated cluster or hub of post-transcriptional regulation, as bidirectional signaling circuits. Recall that the BiLSTM captures patterns by processing sequential gene expression data in both forward and backward directions, enabling it to learn gene signatures that drive past and future states. This is relevant for identifying bidirectional or feedback-regulated mechanisms in complex processes like AML progression. According to GeneCards database, METTL17 (a putative mitochondrial RNA methyltransferase) as well as PARP2, and APEX1, involved in DNA repair and chromatin remodeling are signatures with potential epigenetic roles in AML trajectory prediction. 

The Transformer model (Figure~\ref{fig:aml_dl_importance_heatmaps}) highlights broader metabolic and structural adaptation markers, with HBB, HBA1, HBA2, and TPT1 emerging as key regulators of erythropoiesis and possibly indicating disrupted differentiation due to overlapping erythroid lineage signals. This dsyregulated developmental process is further highlighted by the key signatures of H3F3A/B genes linked to chromatin dynamics and may signal epigenetic reprogramming relevant to AML trajectory. According to GeneCards database, NEAT1 is a long non-coding RNA known to regulate chromatin accessibility, and possibly a signature of the disrupted oncofetal programs. JUND encodes a transcription factor involved in AP-1 signaling, associated with hematopoietic stress responses and leukemic transformation. Hence, the Transformer shows capability of predicting gene signatures relevant to the differentiation control processes of AML trajectories.  EEF1A1, GAPDH, and ACTG1 reflect translational control and cytoskeletal remodeling, while S100A4 and VIM are known for their roles in cancer plasticity and EMT transition. The strong presence of NACA, EIF1, and FAU, some of which were also observed in earlier DEG analyses, suggests translational regulation in developmental patterning. These results indicate that while LSTM and BiLSTM emphasize immune-related differentiation plasticity, the Transformer model captures causal signatures of hematopoietic lineage-specifc regulatory networks, and stalled terminal differentiation critical for AML plasticity, and evolution. All three longitudinal AI algorithms forecasted epigenetic signals predicting AML cell fate trajectories and developmental patterns (attractors) along their gene expression state-space. For the first time, we used these predictive algorithms for identifying causal biomarkers that steer attractor dynamics and govern lineage-specific cell fate decisions in AML.

H3F3A encodes the histone variant H3.3, and mutations in this gene, particularly H3K27M, drive epigenetic dysregulation in paediatric high-grade gliomas (HGG), especially diffuse midline gliomas~\cite{Jessa2019, Saratsis2024, Wang2025}. This driver gene mutation results in the stalled differentiation of these tumors, contributing to their plasticity and aggressive behaviors (poor prognosis)~\cite{Saratsis2024,Wang2025}.

H3F3B encodes histone variant H3.3, just like H3F3A, but they have distinct genomic locations and regulatory mechanisms, although H3F3B mutations are much rarer and less characterized. Our findings suggest they may be epigenetic drivers steering AML differentiation and progression. Furthermore, we observe TPT1 as a progression marker in AML in some of the algorithms. TPT1 plays a key role in glioma plasticity by regulating cell fate commitments and maintaining tumour heterogeneity. Our previous studies identified TPT1 as a master regulator of the transcriptional networks steering plasticity and differentiation dynamics in paediatric glioblastoma~\cite{Uthamacumaran2022,Uthamacumaran2025}.
 
Thus, the emergence of markers such as H3F3A, H3F3B, and TPT1 as critical transition genes in AML progression, identified by transformer networks, suggests an underlying epigenetic plasticity network or mechanism that overlaps with paediatric cancers, such as high-grade gliomas, and contributes to the aggressivity of AML.

\subsection*{NNMF Clustering-based BDM Signatures Reveals Immune Dysregulation, Epigenetic Suppression, and Hematopoietic Cell Fate Biases in AML Trajectory}

We used NNMF clustering-based BDM shift analysis, since NNMF has been shown to effectively reconstruct the developmental trajectories of paediatric high-grade gliomas from single-cell gene expression data~\cite{Wang2025,Jessa2019}. Instead of RNA velocity, we used BDM as a causal discovery tool on this hierarchical clustering. We predicted that our approach would identify latent structure in AML gene expression dynamics and adaptive state-transitions, enabling us to contrast the local, yet longitudinal perturbation patterns captured by the LSTM and Transformer with global (attractor-level) regulatory transitions steering cell fate bifurcations. As a modularity detection algorithm, NNMF captures transcriptional modules (programs), and the higher-order organization and scaling of the network processes, in gene expression patterns. Hence, combined with BDM perturbation, it could decipher the coordinated gain or loss of complexity and causal information dyamics across functional gene modules.

Table~\ref{tab:nnmf_bdm_table} summarizes the top 30 genes with the most extreme BDM shifts derived from NNMF clustering. Genes with strong negative BDM shifts, such as H3F3B, EEF1A1, S100A4, HLA-DRA, HMGB1, NACA, and ACTG1, indicate a coordinated loss of immune regulatory, metabolic, and epigenetic programs. Specifically, H3F3B, suggests epigenetic dysregulation of cell fate decisions. In contrast, genes with positive BDM shifts, including HIST1H4C, S100A9, and IGKC, highlight immune pathway activation, inflammation, and chromatin remodelling. Some of these signatures suggested lymphoid lineage programs as regulators of AML cell fate trajectories. These patterns also support the findings from both the Transformer and LSTM models, which also highlighted disrupted erythroid genes (HBB, HBA1/2, GYPA/B) and lncRNAs (NEAT1), reinforcing that AML fate trajectory is marked by disrupted differentiation processes.

\subsection*{Analysis of Key Overlapping Genes and Nervous System Development in Enrichment Pathways}

Other than the above-discussed neurodevelopmental signatures overlapping with AML progression predictors, such as SALL2, H3F3A/B, KDM5B, and WNT signaling, to name a few, key genes identified in this enrichment analysis via g:Profiler include PCDHA2, PCDHA1, MYOT, HTR2A, GDF6, and others. PCDHA1/2 and related protocadherins play a critical role in homophilic cell adhesion, a key feature in neural circuit formation and synaptic specification. These protocadherins are implicated in neurogenesis, guiding neuronal connections and regulating differentiation of neural stem cells, including neural tube and neural crest cells. Along with other ECM remodelling and cell adhesion regulators identified herein, these findings suggest the role of mechanotransduction and ecological niche construction in the collective behaviors of leukemic cells during AML progression. Further, it suggests that immune-microenvironmental signaling as a cybernetic scaffold of AML fate plasticity. Additionally, GDF6 (Growth Differentiation Factor 6) has an established role in neurodevelopment, particularly in forebrain patterning and neural tube/crest-derived tissue specification.

\subsection*{Transcription Factors in Neurodevelopment and Hematopoietic Lineages}

As shown in Table \ref{tab:gene_enrichment}, using the Bayesian ridge regression algorithm, several transcription factors (TFs) are enriched in this analysis, particularly those known for their role in neural crest and hematopoietic differentiation. These include DLX6, OCT1 (POU2F1), POU3F2, HOX genes (HOXB6, HOXB7, HOXC8, HOXA6, HOXD8), and SOX10. 

- DLX6 is a critical player in forebrain GABAergic interneuron specification.
- POU3F2 (BRN2) is essential for neural progenitor proliferation and differentiation, particularly in the development of the cerebral cortex.
- SOX10 is a master regulator of neural stem cell fate, governing migration and differentiation into peripheral glial cells, melanocytes, and sympathetic neurons.
- HOX genes (HOXA6, HOXB6, HOXB8, HOXC8, HOXD8) are known for their role in segmental identity along the anterior-posterior axis, particularly in specifying neural tube and spinal cord patterning. However, their involvement in hematopoiesis is also well documented, regulating hematopoietic stem cell self-renewal and differentiation into myeloid and lymphoid lineages. The presence of these transcription factors suggests an intersection between neural stem cell-derived structures and hematopoietic lineage specification, pointing to shared developmental pathways in cell lineage determination and cell fate decision-making.

\subsection*{AML Cell Fate Decisions are Shaped by the Brain-Immune-Leukemia Axis and Neuroplastic Signals}

Using Bayesian Ridge regression (BRR), we observed strong evidence of neurotransmission, synaptic signaling, neuroplasticity markers, and neurodevelopmental programs steering AML cell fate trajectories. This algorithm, provides a robust causal inference tool of latent transcriptional drivers guiding lineage dynamics.

BRR complements the AID framework, as a probabilistic linear model with regularization, while AID captures nonlinear, algorithmic complexity and perturbation-sensitive network patterns. Bayesian methods highlight key features with causal associations, while AID detects emergent, dynamic, and non-statistical patterns, reinforcing complexity as an emergent hallmark of AML ecologies and their plastic, evolving cell fate attractors. 

Various genes belonging to the protocadherin alpha (PCDHA) cluster, HTR2A (serotonin receptor), SLC1A6 (glutamate transporter), and GDF6 (growth differentiation factor 6; belonging to the BMP family) are all Bayesian Ridge signatures related to neuronal signaling, neuronal identity, neurotransmission, and synaptic plasticity. For instance, GDF6 might be involved in neural tube and neural crest formation, while the PCDHA are neuronal-specific cell adhesion molecules. The emergence of serotonin receptor and glutamate transporter further indicate that neuronal signaling markers are steering AML cell fate trajectories. 

Furthermore, transcription factors such as DLX6, SOX10, MITF, HOXA6, HOX7, and HOXD1, among other FOX and HOX gene signatures, were significantly associated, reflecting their roles in embryonic development. We interpret this as further evidence for the cross-talk between ectoderm-mesoderm pathways in cell fate (lineage) bifurcations in AML, suggesting that morphogenetic processes involved neural tube and neural crest formation, and pattern formation are dysregulated. These findings reinforce the relevance of neural differentiation-specific regulatory mechanisms in steering AML cell fate trajectories. The neuronal signaling and neurotransmission cues, also support the systems medicine view of AML as complex adaptive ecosystems, and systemic communication disorders (cybernetic systems). These signals may also intersect with hematopoietic (dys)regulation, suggesting a brain–immune–leukemia interface that shapes AML cell fate decisions through neuro-hematopoietic communication. 

\subsection*{Common Lineage Bifurcation in paediatric AML and Neurodevelopment}

Some scholars argue that paediatric AML appear to originate from stem cell-like progenitor cells with retained multilineage potential and developmental plasticity~\cite{boileau2019h3mutant, mumme2023scAML, lambo2023}. In our study, we demonstrate the complexity of this multilineage potential—not as a fixed differentiation hierarchy, but as a decentralized intelligence, or a fluid spectrum of plastic and reprogrammable states. Our analyses reveal an overlap between AML-associated gene expression patterns and neurodevelopmental transcription factors, particularly developmental signatures and plasticity regulators such as HOX gene clusters and SOX10, suggesting a shared embryonic program reactivation akin to that seen in paediatric high-grade glioma (pHGG)~\cite{Larsson2024, Wang2025, Jessa2019}. Neural tube and neural crest cells—derived from the neuroectoderm—mesodermal cross-talk could possibly serve as a key developmental bridge between neuronal and mesenchymal lineages, including plastic hematopoietic derivatives or hybrid (partially committed) states. While neural stem cells (NSCs) and hematopoietic stem cells (HSCs) arise from ectodermal and mesodermal germ layers respectively, they may share developmental cross-talks of transcription factors such as DLX6, POU3F2, and HOX genes. This convergence suggests that some AML cases may emerge from progenitor-like states expressing neural-like programs, as a hallmark of their stalled differentiation and self-renewal. Gene enrichment analyses further support this, highlighting co-regulation between neural and hematopoietic bifurcation points. 
These findings support a disrupted developmental landscape, where embryonic (morphogenetic) patterning processes are hijacked or stalled in leukemogenesis and drive disease progression. We predicted markers that steer these cell fate bifurcations and enable dynamic transitions across lineage trajectories, possibly governed by underlying attractor dynamics that reflect the dysregulation of normal developmental programs.

\section*{Limitations and Mechanistic Interpretations of Cell Fate Cybernetics}

Our systems medicine framework integrates AID, deep learning, as a neurosymbolic approach to dynamical systems modelling and prediction of paediatric AML state-transitions. While RNNs and Transformers captured long-range dependencies and identified critical transition genes, their effectiveness is limited by sparse clinical timepoints—diagnosis (DX), remission (REM), and relapse (REL)—which constrain the resolution of trajectory inference. Though supported by the emergence of epigenetic regulators like \textit{H3F3A/B} and \textit{KDM5B}, multi-omic validation remains a future priority.

Interpretability of deep learning remains a fundamental challenge in systems medicine and Big Data analyses, particularly for causal inference tasks. As Pearl has emphasized, Bayesian updating alone lacks the power to infer causation without structured causal models \cite{pearl2009causality}, and current Large Language Model (LLM)-based approaches are fundamentally statistical rather than causal. Our transformer model, while inspired by LLM architectures, represents a simplified version designed to uncover temporal gene importance.

To overcome some of these limitations, we leveraged BDM perturbation, in combination with feature importance rankings via the input and attention weights of the neural network layers. These algorithms, despite analyzing over 20,000 genes, consistently identified a small subset of top-ranked features (20–30 genes) as highly important features—demonstrating their predictive ability to identify putative causal and mechanistic drivers of AML plasticity with precision, underscoring the biological validity and explainability of the learned models. To further validate their biological explainability, other interpretable causal graph-based deep learning algorithms such as graph-based attention networks, may be explored. AID tools rooted in algorithmic complexity offer a promising alternative by identifying the simplest generative rules underlying transcriptional dynamics and disease progression, as demonstrated herein. 

Our results converge on three regulatory axes—epigenetic dysregulation (e.g., \textit{H3F3A/B}, \textit{NEAT1}, \textit{KDM5B}), immune reprogramming (e.g., \textit{S100A9}), and developmental (morphogenetic) patterning (e.g., \textit{DLX5/6}, \textit{HOX}, \textit{TLE1}, \textit{FOXC1})--suggesting that AML may reflect a stalled differentiation state along the Waddington epigenetic landscape~\cite{Jimenez2023}. These developmental signatures reveal morphogenetic control loss and plasticity, placing leukemic cells near unstable attractor states, stuck from terminal differentiation (stable attractors). Notably, genes like \textit{TPT1}, \textit{KLF4}, and \textit{POU3F2} suggest neurodevelopmental programs are co-opted for stemness and resistance, reinforcing links between gliomas and leukemias.

Mechanistically, overlapping transcription factors and plasticity regulators shared by neural stem cells (NSCs) and hematopoietic stem cells (HSCs)—such as \textit{SOX10}, \textit{DLX6}, and \textit{HOX} clusters—point to a possible bifurcation in early progenitor states, particularly involving neural crest-like intermediates~\cite{Larsson2024, Wang2025, Jessa2019}. This supports the hypothesis that paediatric AML may emerge from stem-like progenitors transiently expressing neural-like features. Consistent with this, Boileau et al. showed that mutant \textit{H3} histones promote pre-leukemic expansion and AML aggressiveness by modulating lineage trajectories~\cite{boileau2019h3mutant}.

While our study focused on top DEGs and high-importance features to derive meaningful plasticity markers regulating cell state-transitions, prospective analyses should include larger, temporally resolved, multi-omic datasets to refine causal mechanisms across multiscale processes. Therapeutically, our results suggest that modulating voltage-gated channels, developmental morphogens, or niche-level bioelectrical signaling and cell adhesion molecules may reprogram leukemic cells away from pathological 'attractors' with malignant traits and maladaptive behaviors, towards stability by restoring differentiation—a teleonomy aligned with patient-centered precision oncology. 

The limitations of single-cell studies include technical biases inherent to scRNA-seq capture technologies, such as dropout events and noise. While we applied log-normalization and statistical methods for differential expression analysis to address batch effects and mitigate some of these challenges, residual biases may remain. One such data-driven limitation lies in the representativeness of the single-cell data, which may not fully capture the diversity of patient populations, comorbidities, or health access and healthcare disparities influencing disease progression. The role of psychosocial factors, ethnic variation and systemic inequities in shaping transcriptional dynamics remains underexplored. Future studies should validate our findings across larger, more inclusive cohorts with time-series gene expression profiles, to ensure equitable, translatable, and excellence-driven advancements in precision medicine and quality patient care.

The identified transcriptomic signatures steering cell fate trajectories may also pave the way for improved non-invasive and precision diagnostics, subtype classification, longitudinal real-time monitoring and risk stratification, treatment response prediction, and liquid biopsy-based real-time monitoring of early disease, progression, or recurrence. As such, we propose integrating multi-omic modalities including single-cell epigenetics, proteomics (e.g., CyTOF mass spectrometry), metabolomics, and lipidomics—to decode causal biomarkers across multiscale cancer processes.

\subsection*{Lineage Bias Toward Hematopoietic Cell Fates and Morphogenetic Blockade in Pediatric AML}

Our study challenges the conventional view of pediatric AML as a disease composed of discrete leukemic stem cell subpopulations undergoing stochastic drift and binary fate bifurcations. Instead, we demonstrate that leukemic cells often emerge from a fluid, hybrid continuum of stem-like progenitor states—primarily common myeloid progenitors (CMPs), but with transcriptional signatures resembling lymphoid-like programs—and become arrested in states of stalled differentiation. These cell states occupy intermediate, plastic, and reprogrammable identities, unable to fully commit to terminal hematopoietic fates. Despite AML’s multilineage plasticity, cell fate (lineage) bias and teleonomic differentiation trajectories were inferred from BDM signatures, with attractor dynamics favoring monocytic (e.g., CD14, S100A8, ARG1), erythroid (e.g., KLF1, RHD, HBB), and megakaryocytic-associated programs.

This biased distribution suggests that leukemic heterogeneity arises not from purely random mutational or selection processes, but from a developmentally constrained attractor dynamics, shaped by early epigenetic priming and regulatory circuit disruptions. Moreover, we identify perturbed bioelectrical and membrane excitability regulators—including KCNH2, KCNE5, and SLC6A8—which are critical for maintaining electrochemical gradients and cell-cell communication during patterning processes. Their dysregulation likely contributes to the destabilization of fate trajectories and supports a model in which AML cell plasticity is actively maintained by disrupted biophysical signaling networks, contributing to maldaptive behavioral patterns and transcriptional heterogeneity.

There appears to be a form of collective intelligence steering AML trajectories through the cell fate decision space, directed toward teleonomic (goal-directed) behaviors and biased lineage identities. These lineage biases can be understood as divergent branches from a shared bifurcation point within the hematopoietic developmental (attractor) landscape. Our results suggest that leukemic cells are arrested near a CMP-like attractor state, with multilineage cell fate bifurcations. Specifically, our findings suggest that leukemic cell fate bifurcations arise predominantly from the megakaryocyte–erythroid (MEP) and granulocyte–monocyte (GMP) branches of the common myeloid progenitor (CMP) state. This places the leukemic cell of origin near a CMP-MEP/GMP bifurcation point, where disrupted signaling and lineage crosstalk distort the attractor landscape, trapping cells in pathological trajectories and a high-plasticity regime characterized by hybrid lineage boundaries, and destabilized differentiation hierarchies. The emergence of IGKC, IGLC2, HLA-DRA, and other immunoglobulin or MHC-related genes—suggest that pediatric AML exhibits lymphoid transcriptional features. This suggests a hybrid progenitor origin or a state of partial lymphoid transcriptional reprogramming, where transdifferentiation toward lymphoid lineages may represent a favored teleonomic endpoint of the cell fate attractor dynamics. However, whether these signatures reflect partial lineage reprogramming along leukemic cell fate trajectories or arise from infiltrating immune populations remains to be elucidated. Conversely, if functionally validated through CRISPR-RNAi screens or causal perturbation assays, this raises the possibility that therapeutically redirecting AML cell fates toward lymphoid transdifferentiation could restore developmental directionality and constrain malignant plasticity.

Furthermore, expression of plasticity regulators such as TPT1, S100A4, VCAN, and chromatin remodelers like KLF1 and H3F3A/B, reinforce Stalled or repressed terminal differentiation programs. The emergence of H3F3A/B in our findings suggests a potential link to KDM6A, a histone demethylase in the same H3K27 methylation–demethylation axis, which is implicated in epigenetic plasticity and developmental lineage bias. Loss-of-function mutations in KDM6A occur in AML, and have been shown to confer malignant traits such as therapy resistance~\cite{stief2020loss}. Furthermore, H3F3A/BB, could possibly interlink AML to oncofetal programs and neural developmental reprogramming, suggesting cross-lineage attractor interference from neuroectodermal axes.

Convergence on erythroid/complement and iron metabolism programs (RHD/CR1L/FCN1; iron/VEGF enrichment) plus ion-channel signatures (such as KCNH2/KCNE5) predicts underexplored interventions: bioelectric modulation, iron–angiogenic coupling, and CD14/TLR feedback blockade as an immunotherapy handle. The iron–VEGF enrichment additionally represents a possible metabolic target for therapeutic intervention.

Several other cell fate identity markers overlap with plasticity signatures identified in pediatric gliomas, implying a possible developmental convergence. This may indicate a shared origin, or initial conditions, rooted in early embryonic programs, involving potential ectoderm and mesoderm cross-communication and stalled morphogenetic processes, contributing to the complex pathogenesis of pediatric AML. Thus, these findings inform causal biomarker discovery and cell fate reprogramming strategies such as precision gene editing, and differentiation therapies, targeting both lineage-specific transcriptional programs and bioelectrical/epigenetic control circuits regulating AML plasticity.

\subsection*{Differentiation Therapies and Cellular Reprogramming Targets}

Based on our AI-driven biomarker discovery and combinatorial perturbation analysis, we propose several testable therapeutic strategies aimed at reprogramming leukemic plasticity in paediatric AML toward stable, differentiated states or reversion to benign-like phenotypes. These combinations leverage both CRISPR/RNAi screens (CRISPR inhibition (CRISPRi)/CRISPR activation (CRISPRa), or RNA interference) and small-molecule or pharmacological perturbations to 'untrap' epigenetic bottlenecks and steer developmental pathways towards more 'constrained' plasticity states, several key targeted therapy combinations have emerged:  

(1) WNT pathway activation (via GSK3$\beta$ inhibition) combined with epigenetic repressors such as KDM1A (LSD1), KDM5B, or ASH1L inhibitors;  
(2) KDM5B inhibition with TLE1 repression, a more target-specific combination;  
(3) Inhibition or deletion of H3F3A, FOXC1, or TPT1, to disrupt oncogenic chromatin states and transcriptional feedback loops;
(4) KLF1 activation or stabilization, potentially via CRISPRa, or BCL11A repression or chromatin-opening agents, to promote erythroid differentiation and suppress leukemic stemness. KLF1 upregulation can be combined with RBP7 (linked to retinoic acid signaling) or retinoic acid, to promote erythroid differentiation.
(5) Co-targeting S100A9 (a key inflammatory mediator with high BDM shifts) or S100A4, and RBP7 to exploit immune-metabolic axes.

KDM1A (LSD1) a histone demethylase that removes mono- and dimethyl groups from H3K4 and H3K9 to maintain adaptive plasticity or stem-like traits in pediatric gliomas\cite{haase2024epigenetic}, while ASH1L deposits H3K36me2, antagonizing Polycomb-mediated silencing. KDM1A, has been implicated in the epigenetic reprogrammability of gliomas and may represent a shared mechanism across pediatric high-grade gliomas and leukemias\cite{haase2024epigenetic}. Our findings suggest these epigenetic targets are implicated in pediatric leukemias, as well. KDM1A and ASH1L were selected based on our combinatorial BDM and Transformer-based feature analyses, as well as the bulk RNA correlation analyses of DEGs, highlighting chromatin regulators like H3F3A/B, KDM5B, and FOXC1 as core drivers of AML plasticity. For instance, ASH1L is functionally linked to H3.3-enriched active chromatin and acts antagonistically to KDM5B, both regulating differentiation programs~\cite{Vann2025ASH1L,Chen2024Chromatin}. Targeting this developmental axis could unblock epigenetic bottlenecks driving the stalled differentiation, and hence, the adaptive plasticity of AML states. Further, we also suggest KDM6A as a possible target based on the emergence of H3F3A/B as a plasticity signature in our algorithmic predictions. Inhibiting KDM5B and activating KDM6A act synergistically on opposite ends of the histone methylation interplay, preserving activating H3K4me3 marks while removing repressive H3K27me3 marks, thereby promoting pro-differentiation gene expression and constraining malignant plasticity.

To validate these, we propose using CRISPR-RNAi screens, alongside available pharmacological perturbation analysis and functional assays. Reprogramming outcomes can be assessed via microscopy (e.g., for protein marker expressions), cell viability assays, and trajectory inference on single-cell RNA-seq data or bulk RNA profiling to monitor transcriptional reprogramming, sub-state dynamics, and the emergence of partially reprogrammed intermediates. These precision targets were selected based on their role in suppressing differentiation and maintaining leukemic stemness. Our computational oncology, and systems medicine framework opens new pathways for predictive biomarkers in patient care, and causal discovery in cell fate engineering and reprogramming strategies as safer, targeted therapies in precision medicine. 

\begin{table}[H]
\centering
\caption{Proposed Cell Fate Reprogramming Combinations as Targeted Therapies}
\label{tab:reprogramming_targets}
\begin{tabular}{p{4cm}|p{3cm}|p{6.5cm}}
Combo / Target & Method & Purpose / Role \\
\hline
TLE inhibition or WNT activation (GSK3$\beta$i) + KDM1A inhibition & Small molecule & Epigenetic unlocking of H3K4 demethylation; promotes myeloid differentiation \\
\hline
TLE inhibition or WNT activation (GSK3$\beta$i) + ASH1L inhibition & Small molecule & Blocks HOXA-driven stemness; WNT activation promotes lineage commitment \\
\hline
WNT activation (GSK3$\beta$i) + KDM5B inhibition & Small molecule & Suppresses H3K4me3 demethylation; de-represses pro-differentiation genes \\
\hline
KDM5B inhibition + TLE1 inhibition & CRISPRi / shRNA / small molecule & Target-specific epigenetic release + derepression of WNT/$\beta$-catenin pathway \\
\hline
ASH1L$\downarrow$ + TLE1$\downarrow$ (+ GSK3$\beta$i) & CRISPRi / shRNA + small molecule & Epigenetic unlock + $\beta$-catenin signaling reinforcement for AML reprogramming \\
\hline
H3F3A/B inhibition & CRISPRi / pharmacological & Destabilizes histone variant control; weakens oncogenic memory \\
\hline
KDM6A activation + H3F3A/B modulation & CRISPRa / small molecule & The emergence of H3F3A/B suggests KDM6A could regulate developmental plasticity in AML aggressivity and progression \\
\hline
TPT1 deletion & CRISPR knockout / pharmacological & Suppresses tumor survival adaptation; sensitizes to differentiation stimuli \\
\hline
FOXC1 repression & CRISPRi & Silences self-renewal program; enhances differentiation competency \\
\hline
KLF1 activation + RBP7 upregulation or Retinoic Acid & CRISPRa / Small molecule (e.g., ATRA) & Induces erythroid differentiation via transcriptional reprogramming and retinoid signaling axis \\
\hline
ARHGEF7, RAB20, RASA3 repression & CRISPRi / CRISPR knockout & Disrupts Rac/Ras GTPase signaling; reduces leukemic plasticity and therapy resistance \\
\hline
S100A9 or S100A4 repression + RBP7 activation & CRISPRi / small molecule & Targets inflammatory-metabolic axes to reduce leukemic plasticity and promote lineage stabilization \\
\end{tabular}
\end{table}

\section{Conclusions}

To conclude, our complex systems framework revealed causal drivers of developmental plasticity and state transitions in pediatric AML. These algorithms provide explainable AI for decoding plasticity markers and the causal processes shaping AML eco-evolutionary dynamics, enabling the prediction of cell fate behavior patterns. For the first time, we combine powerful RNN architectures with algorithmic complexity measures to identify critical plasticity signatures that steer AML progression in gene expression state space, offering predictive biomarkers for cell fate reprogramming and precision medicine. Our predictive algorithms identified causal plasticity biomarkers and network signatures forecasting cell fate trajectories. The observed dysregulated developmental programs suggest that AML navigates pathological attractors, consistent with stalled differentiation and disrupted cell fate (lineage) identity.

Across integrated algorithms, we identified H3F3A/B, KDM5B, S100A9, KCNE5, FOXC1, TLE1, TPT1, GDF6, PCHD1/2, and NEAT1, among other developmental signatures and plasticity markers, as critical regulators of AML state-transitions toward aggressiveness. In specific, our analyses suggest that paediatric AML plasticity is driven by key epigenetic regulators—such as H3F3A/B, KDM5B, TPT1, and TLE1—which are also implicated in paediatric high-grade gliomas (pHGG), suggesting a shared developmental program across these lethal diseases. Bulk RNA correlation and deep learning models converged on these genes as central to chromatin remodeling and cell fate reprogramming. BDM perturbation analysis further uncovered ion-channel regulators (SLC6A8, KCNE5, S100A9, etc.) as key contributors to the bioelectric signaling network shaping AML's patterns of behavior. These findings support the hypothesis that AML cells, through disrupted voltage and signaling dynamics, decouple from hierarchical tissue control, enabling invasive and therapy-resistant phenotypes~\cite{Levin2023}. 

Transcription factors such as DLX5/6, FOXM1, POU3F2, and HOX clusters identified by Bayesian Ridge Regression and Transformer analysis align with both hematopoietic and neural stem cell (NSC) developmental axes, highlighting a convergence of regulatory programs between the nervous system and hematopoiesis. This suggests a bifurcation point in early progenitor differentiation—consistent with epigenetic attractor models—where developmental plasticity is hijacked during leukemogenesis. Although most of the identified signatures are general hematopoietic lineage markers, some such as IGKC and IGLC2, are notable lymphoid-associated markers. This suggest that a lymphoid-like program or hybrid lineage-identity may be activated or retained in the AML leukemic population. This may further support the stalled differenitation and mixed-lineage plasticity in the paediatric AML. Our findings also reveal neurodevelopmental signatures steering AML cell fate bifurcations, enriched with neuroplasticity and neurotransmission signals as possible drivers of disrupted morphogenetic and lineage decision processes in leukemogenesis. These findings also further support the ectoderm–mesoderm cross-talk and a systemic brain–immune–hematopoietic axis underlying disrupted developmental trajectories. 

Our graph network-based complexity signatures, which capture underlying plasticity markers (bifurcation signatures) driving glioma invasiveness and evolution, may also function as predictive biomarkers forecasting disease trajectory. We propose their potential clinical utility as actionable targets for precision medicine, as well as prognostic indicators and early recurrence detection, enabling real-time patient monitoring through integrative multi-omics and liquid biopsy approaches. Hence, our findings support translational opportunities in preventive care and predictive medicine.

Collectively, our findings suggest AML ecosystems are disorders of cell fate decisions, and lineage identity, wherein phenotypic plasticity markers signal stalled differentiation. Our algorithmic predictions highlight the potential of cellular reprogramming-based therapies targeting plasticity markers steering oncofetal developmental programs to restore differentiation, constrain plasticity, and reverse malignant cell states toward stable cellular identities~\cite{Proietti2022,Bizzarri2020,Gong2025}.  Our integration of neurosymbolic AI, via AID, and deep learning enables the forecasting of cell fate decision dynamics and causal biomarker discovery, with translational potential for early recurrence detection, relapse prediction, personalised risk stratification, and precision therapy development in the treatment and care of paediatric AML. 

\section*{Code and Data Availability}
All codes, including our deep learning algorithms (Transformer, LSTM, BiLSTM) and complexity-based network perturbation analyses pipelines, are publicly available at \url{https://github.com/Abicumaran/Longitudinal_AML} for reproducibility, accessibility, and translational medicine.

\printbibliography

\clearpage
\appendix

\section{Phenotypic Plasticity and the Need for Predictive Algorithms to Decode Cell Fate Cybernetics}

Pediatric acute myeloid leukemia (pAML) remains a lethal hematological malignancy and an evolutionary disorder driven by maladaptive behaviors such as clonal adaptation, therapy resistance, and relapse. Its high mortality stems from the interplay of developmental lineage blockade, and microenvironmental remodeling, which enable the disease to navigate and exploit an adaptive evolutionary landscape. Unconstrained phenotypic plasticity serves as the cognitive engine of these malignant traits.

Phenotypic plasticity in pAML resembles a reaction norm, a dynamic continuum bridging differentiation and intratumoral heterogeneity. This plasticity spans from stem-like to invasive phenotypes, and identifying its causal drivers or markers may uncover reprogrammable targets for forecasting and directing cell fate decisions. Many hallmarks of disease progression, such as aggressivity, relapse, therapy resistance, and poor risk stratification, can be traced to this unconstrained plasticity, which fuels developmental instability and adaptive reprogramming.

The leukemic developmental (attractor) landscape can be conceptualized as a distorted Waddington-like attractor space, where disrupted lineage bifurcations and stalled differentiation generate pathological attractor states and maladaptive behaviors. However, current models fall short of capturing such complex landscapes, thus limiting their utility for predictive medicine and early relapse detection.

To address this gap, robust causal inference algorithms are needed, capable of identifying critical “tipping points” along the gene expression state space, where malignant states may revert (reprogram) to quiescent or normal-like behaviors. These phenotypic transitions are governed by poised, metastable chromatin states sensitive to developmental timing, morphogen gradients, and regulatory network dynamics, i.e., core features of AML plasticity.

Thus, we adopted complex systems theory approaches to forecast and identify causal drivers (biomarkers) of developmental plasticity and leukemic state transitions in Pediatric AML (pAML) transcriptomics. By applying complexity science—combining nonlinear dynamics, network theory, and causal inference—we reconstruct the plasticity network dynamics that govern cell fate decisions, capturing transitional states overlooked by conventional reductionist methods and proposing actionable targets for differentiation-based reprogramming therapies. In this study, we combined for the first time, time-series based deep learning algorithms, including Recurrent Neural Networks and Transformers with algorithmic information dynamics (AID), and causal network perturbation analysis on longitudinal, patient-matched single-cell transcriptomics at three time-points (diagnosis, remission, relapse) to reconstruct attractor geometries and forecast (critical) tipping-points of fate bifurcations, as plasticity regulators. 

Our findings reveal that cytoskeletal-matrix remodeling, immune-metabolic reprogramming, oncofetal (morphogenetic) programs, and neuronal-like lineage biases are tightly coupled to the AML developmental attractor landscape. This systems medicine framework, grounded in complexity science, bridges computational theory, developmental biology, and translational therapeutics, to advance precision oncology strategies for understanding, predicting, and treating aggressive hematological malignancies, like pAML. 

\section{Background}

Deep learning has become a cornerstone in modeling dynamical systems, particularly in single-cell multiomics, where understanding cellular fate transitions is crucial. Recent AI-driven approaches leverage transformers, RNNs, and graph-based deep learning for cell-type annotation, clustering, and trajectory inference. Graph-based neural networks (GNNs) such as Graph convolutional networks (GCNs), and graph autoencoders (GAEs), utilize attention-based mechanisms to refine feature extraction from single-cell transcriptomics~\cite{liu2024sclega,li2023attention, Szalata2024}. Transformers have also been deployed for cell-type annotation, with models such as scBERT capturing hierarchical dependencies across single-cell RNA-seq data~\cite{chen2023transformer,lan2024transformer}. Furthermore, transformer-based gene regulatory network (GRN) inference models, such as STGRNS, provide insights into the latent gene interactions in cellular differentiation~\cite{xu2023stgrns}. Moreover, LSTM-based models such as ScLSTM leverage recurrent memory mechanisms to capture temporal and relational patterns in time-series scRNA-seq data, enabling accurate detection of cell types and differentiation hierarchies by modelling similarity in gene expression across cells~\cite{jiang2023, Sadria2024}. Building on this, FateNet integrates deep learning with dynamical systems theory to identify bifurcation points in cell fate trajectories, using LSTM and convolutional neural networks (CNN) to predict cell fate transitions and the emergence of cellular identity during differentiation processes~\cite{Sadria2024}.

Traditional approaches for modelling cellular dynamics, such as single-cell fate decision-making rely on pseudotemporal ordering, manifold learning, and optimal transport theory. The Waddington-OT algorithm reconstructs probabilistic trajectories through optimal transport to infer cellular differentiation dynamics~\cite{schiebinger2019optimal}. Similarly, TrajectoryNet uses dynamic optimal transport to model cellular differentiation as a continuous trajectory~\cite{tong2020trajectorynet}. However, these methods assume an underlying statistical distribution, which may not fully capture the emergent nonlinear attractors governing single-cell disease progression. Alternative deep learning methods, such as LSTMs, have demonstrated potential in hierarchical clustering for single-cell phenotype detection~\cite{jiang2023}, yet remain underexplored in cancer progression modelling. Reinforcement learning has also been introduced for single-cell fate choice mapping, offering a promising framework for adaptive decision-making in complex systems~\cite{fu2024reinforcement}. While deep learning-based time-series models such as temporal attention networks and LSTMs have been used for gene expression forecasting~\cite{yuan2021deep}, they lack causal inference. 

Current approaches fail to establish causality in cellular state transitions, necessitating the integration of AID. A robust dynamical complexity measure, i.e., one that quantifies the causal information structure and predicts the temporal evolution of complex adaptive systems, such as gene regulatory networks, is lacking in traditional analyses, which typically rely on pseudotime ordering, entropy, or probabilistic surrogates~\cite{Zenil2018, Zenil2019,zenil2019algorithmic}. Unlike these suboptimal measures, AID quantifies the algorithmic generative rules underlying a complex dynamical system's behavior, inferring causal network complexity~\cite{Zenil2018, Zenil2019,zenil2019algorithmic}. Such graph-theoretic complexity could predict cell fate transitions and developmental trajectories, serving as robust plasticity biomarkers in disease progression~\cite{Uthamacumaran2022,uthamacumaranzenil2022, Uthamacumaran2025}. AID, through perturbation-based complexity analysis, allows causal inference of predictive signatures driving AML progression, and plasticity (state-transitions), without requiring prior statistical assumptions. By integrating AID with deep learning-based longitudinal predictive modelling, we present a novel approach for inferring AML progression as a complex adaptive dynamical system.

For instance, recent studies in data-driven systems biology framework have modelled the core gene regulatory network (GRN) controlling leukemogenesis in IDH-mutant AML, revealing mechanisms of impaired differentiation and cell state transitions using network perturbations~\cite{Katebi2024}. Although the study identified perturbations that destabilise the state of AML and suggest therapeutic targets, it did not capture causal predictors or temporal state transitions toward aggression, highlighting the need for time-series gene expression data integrated with causal discovery algorithms~\cite{Katebi2024}. Network dynamics and attractor inference are some tools for capturing goal-directed (teleonomic) behaviors in the evolutionary trajectories of cancer ecologies. However, the lack of a robust graph complexity algorithm for causal pattern discovery and prediction has limited this avenue for the most part of systems oncology.

Therefore, we predicted that by combining systems medicine principles, such as network science, dynamical systems theory, and algorithmic information theory, we can decipher the goal-directed, symbolic communication, cybernetic feedback loops, and semiotic information flows embedded in AML progression. These processes regulate cell fate decisions, behavioural patterns, and transition dynamics toward aggressive cancer phenotypes. The GRN model reveals how cancer ecosystems evolve as self-regulating, cognitively adaptive systems modulated through (nonlinear) attractor dynamics and perturbation-based reprogramming~\cite{Katebi2024, Gong2025}. To forecast these emergent patterns of behavior (attractors) during cell fate transitions, we combined causal discovery (AID) with RNNs and deep learning algorithms for predictive modeling of cell fate decisions. Compared to traditional pseudotemporal methods, our symbolic AI-driven framework provides causal inference to predict critical plasticity markers driving malignant traits such as AML invasion and therapy resistance. 

Hence, in this study, we employed symbolic AI methods such as AID, i.e., perturbation analysis with BDM and RNNs to identify critical transition genes and biosignatures influencing AML progression from single-cell gene expression matrices (scRNA-Seq)~\cite{zenil2019algorithmic}. We used log fold-change and p-value filtering from statistical AI approaches to compute differentially expressed genes, complemented by deep learning models like LSTM networks and transformers to capture long-range temporal dynamics and extract feature importance from the network layers' weights. These methods reveal developmental hierarchies and differentiation trajectories by analyzing how gene expression profiles shift between attractor states along a longitudinal trajectory from diagnosis (DX), to remission (REM) and relapse (REL). This dynamical systems approach provides insights into the underlying regulatory networks driving relapse dynamics. These methods jointly revealed state-transition genes driving changes from diagnosis to relapse, highlighting key regulators and developmental patterns of AML evolution. Judea Pearl's work on causality emphasizes that understanding the cause-and-effect relationships requires going beyond correlations to model the underlying mechanisms and interventions driving the observed data. Hence, we compared the predictive power of RNNs combined with AID to a baseline of Bayesian inference and correlation metrics.

Deep learning models like RNNs capture complex temporal dependencies and non-linear relationships in sequential data, used for inferring causal discovery in dynamic systems. Compared to traditional Bayesian inference, which relies on explicit priors and interpretable probabilistic frameworks, RNNs excel in high-dimensional, noisy data but often lack transparency in causal reasoning. To overcome this black box, we used feature importance from the learned weights to render them as explainable AI and identify 'plasticity signatures' and network biomarkers capturing AML state-transition dynamics, as predictive features of AML progression. In contrast, AID utilizes algorithmic complexity to directly measure causal structure, avoiding the heavy training and parameter optimization needed for RNNs, and making it computationally efficient for small or sparse datasets.

On the other hand, AID, grounded in measures like Block Decomposition Method (BDM) and BDM-based network perturbation analysis, extends beyond traditional information theoretics such as Shannon entropy by quantifying the algorithmic complexity of complex adaptive systems, capturing causal patterns or randomness in gene expression networks (matrices)~\cite{uthamacumaranzenil2022}. Unlike entropy, which reflects average uncertainty, BDM estimates the minimal generative rules underlying the complex system's behavior, allowing the discovery of latent causal mechanisms. When combined with perturbation analysis in silico, AID enables robust causal discovery by identifying which genes or subnetworks govern network topology, stability, and causal dynamics. Our recent work  demonstrates that AID can reconstruct attractor landscapes and infer causal perturbation hierarchies by identifying algorithmically sensitive genes or modules within complex gene regulatory networks~\cite{Zenil2018, Zenil2019,zenil2019algorithmic}. AID-driven perturbation analysis revealed hidden structural regularities, phase transitions, and critical tipping points within these attractors, which are not captured by traditional information theoretics, or pseudotime-based trajectory inference algorithms~\cite{Zenil2018, Zenil2019,zenil2019algorithmic}. Therefore, AID provides a powerful causal discovery framework for how network topology and modularity encode collective, maladaptive behaviors such as plasticity, and resilience in complex adaptive systems like cancer. By identifying causally influential genes and perturbation-sentiive links, nodes, or modules within the longitudinal gene expression networks, AID enables targeted reprogramming of cancer cell types toward stable, non-malignant attractor states~\cite{zenil2019algorithmic, uthamacumaranzenil2022, Uthamacumaran2022,Uthamacumaran2025}.

AID's potential remains largely underexplored in cancer research, especially in decoding cellular cybernetics such as forecasting differentiation dynamics (cell fate decision landscapes) and identifying their predictive signatures (i.e., plasticity biomarkers). To the best of our knowledge, this is the first empirical study integrating RNNs (LSTM, Transformers) with AID to decode AML state-transitions, paving the way for reprogramming strategies like differentiation or cancer reversion therapy to control and stabilize aggressiveness~\cite{zenil2019algorithmic, Zenil2019,Zenil2018}. Thereby, our findings advance computational oncology and systems medicine by combining RNNs and AID-driven causal pattern discovery to map and infer cell fate decision dynamics in cancer ecosystems.

\section{Systemic Communication Breakdown and Cell Fate Identity Disorders as Hallmarks of AML Ecology}

Our findings reveal a neurodevelopmental and brain–immune–hematopoietic axis in pediatric AML, with transcriptional signatures linked to neuroplasticity, neurotransmission, and immune signaling co-occurring alongside hematopoietic lineage biases and morphogenetic markers. This reveals a plastic spectrum of cell fate identities, where leukemic cell fates exhibit biased differentiation trajectories, as if navigating a complex, unstable attractor, rather than discrete, fixed subpopulations. Such cell fate bifurcations reflect a breakdown in systemic communication—what would normally be coordinated, developmentally regulated pattern formation (morphogenesis) becomes dysregulated. Cell fate decisions, typically orchestrated through rules and constraints, signaling feedback loops, and differentiation hierarchies, now exhibit cell fate identity disorders, suggesting a fragmentation or deviance of collective cellular intelligence, with plastic (emergent) social-like constraints. These observations support two key models, under the complex systems framework:

1. AML as a Complex Adaptive Ecosystem: The presence of neurotransmitter signaling, immune–metabolic programs, lymphoid transcriptional features, and microenvironmental remodeling signals—including cytoskeletal and extracellular matrix cues, bioelectric signals, ion channels, and mechanotransduction (e.g., cadherins)—supports the view of AML as an ecological system, with complex leukemia-microenvironmental interactions and niche construction. Here, cell identities and maladaptive patterns of behaviors are dynamically shaped by their systemic “umwelt.” The miscoordination among these signals leads to cellular dissociation and fragmented identities, reflecting disrupted morphogenesis and hence, ecological dysregulation.

2. AML as a Complex Cybernetic System: The fragmentation and dissociation of cell identity, and malignant plasticity can be interpreted as emergent, behavioral patterns of a maladaptive cybernetic response to disrupted information flow during pattern formation, and nice construction. In this framework, cancer arises from miscommunication across regulatory networks and loss of coherent feedback loops. Cybernetics, the processes of control, communication, and feedback within complex systems, offers a lens to understand AML’s aberrant signaling dynamics, and plasticity as both vulnerability and maladaptive behaviors. Through this lens of collective intelligence, AML cell fate decision-making, are guided by fate plasticity as a cybernetic scaffold, creating fluid-like, and dissociative identities through nonlinear and reprogrammable trajectories. As such, we can define this adaptive intelligence as creativity, or creative dynamics. 

AML's Achille's heel also reveals a therapeutic window: plasticity implies reprogrammability. By identifying causal biomarkers of cell fate bifurcations and morphogenetic miscommunication, we can tailor precision medicine approaches—such as differentiation therapy or cancer reversion strategies, to restore cellular identity, constrain plasticity, and redirect malignant cell fates toward stable, teleonomic, and functional developmental states. 

3. AML Plasticity as a Cognitive Engine Steering Maladaptive Behaviors: Lastly, our findings underscore complexity not merely as a byproduct of pathogenesis, but as a defining hallmark of leukemogenesis, and cancer progression. Emergent behaviors, nonlinear dynamics, phase transitions (bifurcations), and multiscale interactions constitute the architecture (self-organizing logic) and evolvability engine of the leukemic umwelt, steering identity formation and fate plasticity (reprogrammability).

\begin{table}[H]
\centering
\begin{tabular}{l|l}
\hline
\textbf{Term Name} & \textbf{Adjusted p-value} \\

Calcium ion binding & 1.83E-09 \\
Cation binding & 3.75E-06 \\
Metal ion binding & 4.09E-05 \\
Ion binding & 0.000493 \\
Small molecule binding & 0.000813 \\
Homophilic cell adhesion & 1.26E-18 \\
Cell-cell adhesion & 1.45E-09 \\
Cell adhesion & 3.53E-07 \\
Nervous system development & 0.00013 \\
System development & 0.01886 \\
Plasma membrane & 2.78E-05 \\
Cell periphery & 9.39E-05 \\
Membrane & 0.009843 \\
Eicosanoids & 0.002156 \\
Fatty acids & 0.003428 \\
TF: ZNF2 & 5.88E-12 \\
TF: Oct-1 & 1.30E-09 \\
TF: FoXM1 & 0.000615 \\
TF: CDP & 0.000691 \\
TF: OCT-2 & 0.001292 \\
TF: MITF & 0.003007 \\
TF: POU3F2 & 0.003154 \\
TF: HSF1 & 0.004898 \\
TF: HOXD8 & 0.00833 \\
TF: HOXB7 & 0.01373 \\
TF: DLX-6 & 0.01707 \\
TF: HOXB6 & 0.02219 \\
TF: Sox-10 & 0.0231 \\
TF: HOXB8 & 0.02512 \\
TF: HOXA6 & 0.03644 \\

\end{tabular}
\caption{g:Profiler Gene set enrichment of Bayesian Ridge Regression shown in Figure 3G (main Text). Enriched terms are shown with their respective adjusted p-values.}
\label{tab:gene_enrichment}
\end{table}

\end{document}